\documentclass[a4paper,fleqn,usenatbib]{mnras}

\usepackage[utf8]{inputenc} 
\usepackage{ae,aecompl}

\usepackage{graphicx}	
\usepackage{amsmath}	
\usepackage{amssymb}	

\def\beq{\begin{equation}}
\def\eeq{\end{equation}}
\def\bea{\begin{eqnarray}}
\def\eea{\end{eqnarray}}

\newcommand{\RN}{Reissner--Nordström}

\title[Charge of the Galactic centre black hole]{On the charge of the Galactic centre black hole}

\author[M. Zaja\v{c}ek et al.]{
Michal Zaja\v{c}ek$^{1,2,3}$\thanks{E-mail: zajacek@ph1.uni-koeln.de},
Arman Tursunov$^{4}$,
Andreas Eckart$^{2,1}$,
 and
Silke Britzen$^{1}$
\\
$^{1}$Max-Planck-Institut f\"ur Radioastronomie (MPIfR), Auf dem H\"ugel 69, D-53121 Bonn, Germany\\
$^{2}$I. Physikalisches Institut der Universit\"at zu K\"oln, Z\"ulpicher Strasse 77, D-50937 K\"oln, Germany\\
$^{3}$Astronomical Institute, Academy of Sciences, Bo\v{c}n\'{\i}~II 1401, CZ-14131~Prague, Czech Republic\\
$^{4}$Institute of Physics and Research Centre of Theoretical Physics and Astrophysics, Faculty of Philosophy and Science,\\Silesian University in Opava, Bezru\v{c}ovo n\'{a}m.13, CZ-74601 Opava, Czech Republic
}

\date{Accepted 2018 August 05. Received 2018 May 29; in original form 2017 September 29}

\pubyear{2017}

\begin{document}
\label{firstpage}
\pagerange{\pageref{firstpage}--\pageref{lastpage}}
\maketitle

\begin{abstract}
The Galactic centre supermassive black hole (SMBH), in sharp contrast with its complex environment, is characterized by only three classical parameters -- mass, spin, and electric charge. Its charge is poorly constrained. It is, however, usually assumed to be zero because of neutralization due to the presence of plasma.  We revisit the question of the SMBH charge and put realistic limits on its value, timescales of charging and discharging, and observable consequences of the potential, small charge associated with the Galactic centre black hole. The electric charge due to classical arguments based on the mass difference between protons and electrons is $\lesssim 10^9\,{\rm C}$ and is of a transient nature on the viscous time-scale. However, the rotation of a black hole in magnetic field generates electric field due to the twisting of magnetic field lines. This electric field can be associated with induced charge, for which we estimate an upper limit of $\lesssim 10^{15}\,{\rm C}$. Moreover, this charge is most likely positive due to an expected alignment between the magnetic field and the black-hole spin. Even a small charge of this order significantly shifts the position of the innermost stable circular orbit (ISCO) of charged particles. In addition, we propose a novel observational test based on the presence of the bremsstrahlung surface brightness decrease, which is more sensitive for smaller unshielded electric charges than the black-hole shadow size. Based on this test, the current upper observational limit on the charge of Sgr~A* is $\lesssim 3\times 10^{8}\,{\rm C}$.      
\end{abstract}




\section{Introduction}

The observations of the Galactic centre across the electromagnetic spectrum, ranging from radio to gamma wavelengths, revealed the complex structure of the Nuclear Star Cluster (NSC) as well as that of the gaseous-dusty medium of the central parsec \citep{2005bhcm.book.....E,2007gsbh.book.....M,2010RvMP...82.3121G,2017FoPh...47..553E}. The presence of the concentrated, dark mass at the dynamical centre of the NSC was revealed by the near-infrared observations of stars using adaptive optics. The first proof for the compact dark single object  in the Galactic centre came with the detection and the analysis of the first proper motion of stars orbiting Sgr~A* inside $1'' \sim 0.04\,\rm{pc}$, so-called S stars \citep{1996Natur.383..415E,1997MNRAS.284..576E,1998ApJ...509..678G, 2018arXiv180411014Z}. Based on these and follow-up observations \citep{2016ApJ...830...17B,2017ApJ...845...22P,2009ApJ...692.1075G,2017ApJ...837...30G,2018A&A...615L..15G}, the large mass of the dark object has been confirmed and a more precise value has been determined --  $\sim (4.15 \pm 0.13 \pm 0.57)\times 10^6\,M_{\odot}$ \citep{2017ApJ...845...22P}. If we associate this dark mass with a non-rotating black hole for simplicity, this yields a Schwarzschild radius of $R_{\rm Schw}=1.2 \times 10^{12}\,{\rm cm}(M_{\bullet}/4\times 10^6\,M_{\odot})$ and the expected mean density is,

\begin{equation}
  \rho_{\bullet}=1.7 \times 10^{25} \left(\frac{M_{\bullet}}{4\times 10^6 \, M_{\odot}}\right) \left(\frac{R_{\rm Schw}}{3.9 \times 10^{-7}\,{\rm pc}} \right)^{-3}\,M_{\odot}{\rm pc^{-3}}\,.
  \label{eq_density_blackhole}
\end{equation}

In case of stellar orbits, the tightest constraint for the density of the dark mass comes from the monitoring of B-type star S2 with the pericentre distance of $r_{\rm P} \simeq 5.8 \times 10^{-4}\,{\rm pc}$ \citep{2017ApJ...845...22P,2002Natur.419..694S,2009ApJ...692.1075G,2017ApJ...837...30G,2018A&A...615L..15G}

\begin{equation}
  \rho_{\rm S2}=5.2 \times 10^{15} \left(\frac{M_{\bullet}}{4.3 \times 10^6\,M_{\odot}}\right)\left( \frac{r_{\rm P}}{5.8 \times 10^{-4}\,{\rm pc}} \right)^{-3}\,M_{\odot}{\rm pc^{-3}}\,.
  \label{eq_density_blackholeS2}
\end{equation}
The most stringent density constraint was given by 3$\sigma$ VLBI source size of $\sim 37\mu {\rm as}$ \citep{2008Natur.455...78D,2018ApJ...859...60L}. When combined with the lower limit on the mass $M_{\rm SgrA*} \gtrsim 4 \times 10^5\,M_{\odot}$ based on the proper motion measurements \citep{2004ApJ...616..872R}, VLBI yields the lower limit of $\rho_{\rm SgrA*} \geq 9.3\times 10^{22}\,M_{\odot}{\rm pc^{-3}}$. This is about two orders of magnitude less than the density expected for a black hole of $\sim 4\times 10^6\,M_{\odot}$, see Eq.~\eqref{eq_density_blackhole}. The most plausible stable configuration that can explain such a large concentration of mass emerges within the framework of general relativity: a singularity surrounded by an event horizon -- a black hole, ruling out most of the alternatives \citep{2017FoPh...47..553E}.  

According to the uniqueness or the general relativistic ``no-hair" theorem \citep{1996bhut.book.....H}, any stationary black hole is fully characterized by only three classical and externally observable quantities: mass $M_{\bullet}$, angular momentum $J_{\bullet}$ (often the quantity $a_{\bullet}=J_{\bullet}/M_{\bullet}c$ is used which has a dimension of length), and the electric charge $Q_{\bullet}$\footnote{In case a magnetic monopole could exist, it could be the forth parameter.}.
Thanks to the high-precision observations of stars in the Nuclear Star Cluster, including the innermost S cluster, the current value for the SMBH mass is $M_{\bullet}=(4.3 \pm 0.3) \times 10^6\,M_{\odot}$ \citep{2017FoPh...47..553E}, which is based on different methods, primarily the orbits of S stars \citep{2017ApJ...845...22P}, the Jeans modelling of the properties of the NSC \citep{2013ApJ...779L...6D}, and the general relativistic fits to the double-peaked X-ray flares that show signs of gravitational lensing \citep{2017MNRAS.472.4422K}.  The constraints for the spin $J_{\bullet}$ were inferred indirectly based on the variable total and polarized NIR emission \citep{2006A&A...455....1E}. The spin can be determined based on the modelling of spin-dependent quantities, mainly the light curves of a hot spot or a jet base. In this way, \citet{2006A&A...460...15M} obtained constraints for the spin, which are rather weak and the spin parameter is $a_{\bullet} \gtrsim 0.4$, as well as the inclination, which is inferred based on the stable polarization angle of the flares and tends to be rather large $i \gtrsim 35^{\circ}$. The value of the spin parameter determined based on quasi-periodic oscillations for Sgr~A* reaches a unique value of $\approx 0.44$ \citep{2010MNRAS.403L..74K}, which is consistent with the value inferred from the fitting of the NIR flares.

In general, the charge of the black hole $Q_{\bullet}$ is often set equal to zero due to the presence of plasma around astrophysical black holes. However, a black hole can acquire primordial charge because it was formed by a collapse of a charged (compact) star \citep{2003PhRvD..68h4004R}. It is not clear on which timescales such a charged black hole discharges or alternatively, can increase its charge. Also, from an astrophysical point of view, it is of a general interest if a charged black hole can be observationally distinguished from a non-charged case, clearly depending on the value of the charge.

In addition, electric charge can be loaded or induced by black hole due to its rotation in external magnetic field within the mechanism similar to the Faraday unipolar generator. Such a mechanism is more relevant for supermassive black holes in the local Universe, since the primordial charge information is expected to be lost. The induction mechanism works in such a way that the rotation of a black hole generates electric potential between horizon and infinity which leads to the process of selective accretion of charged particles of plasma surrounding the black hole.
In particular, a rotating black hole embedded in a uniform, aligned magnetic field will acquire an electric charge until an equilibrium value is reached $Q_{\bullet, W}=2B_{\rm 0} J_{\bullet}$, a so-called Wald charge \citep{1974PhRvD..10.1680W}, where $B_0$ is an asymptotic magnetic field strength. There is an evidence that significant and highly aligned magnetic field must be present in the Galactic center with equipartition strength of $10\,{\rm G}$ in the vicinity of the event horizon of the SMBH \citep{2012A&A...537A..52E,2013Natur.501..391E,2015llg..book..391M}. 

The twisting of magnetic field lines threading the horizon of rotating black hole produces an electric field which accelerates the charged particles along the magnetic field lines. Moreover, magnetic field plays the role of a catalyzing element that allows the extraction of rotational energy from rotating black hole through interaction of charged particles with an induced electric field in such processes as the Blandford-Znajek mechanism \citep{1977MNRAS.179..433B} and the magnetic Penrose process \citep{1985ApJ...290...12W}. Both of these processes that allow the energy extraction from rotating black holes require the presence of an induced electric field \citep{2018MNRAS.478L..89D}.


Even a small charge associated with the black hole can have considerable effects on the electromagnetic processes in its vicinity, such as the bremsstrahlung emission and the motion of charged particles as we will show.  
 The value of this small electric charge for black holes embedded in plasma will be necessarily temporary and fluctuating, mainly due to the attraction of oppositely charged particles and/or the variability of the magnetic field in which the black hole is immersed. Even for an extreme case of a charged black hole in vacuum, a spontaneous loss of charge would occur due to pair production with an exponential time-dependency \citep{1975CMaPh..44..245G}.  

In this paper, we revisit the question of a charge, mainly of an electric origin, associated with the Galactic centre SMBH. Previously, several theoretical studies have focused on the spacetime structure of charged black holes \citep{1991JPhy1...1.1005K,1991JMP....32..714K,2011PhRvD..83j4052P,2011PhRvD..84h4002K}. Here we are aiming at the connection between the current theoretical knowledge with a real astrophysical case -- Sgr~A* supermassive black hole, for which we gathered most constraints on its nearby plasma environment \citep{2017FoPh...47..553E} -- in order to put realistic constraints on electric charge of our nearest supermassive black hole. 

 The study is structured as follows. In Section~\ref{section_prospects} we analyse the potential for charging given the plasma properties in the surroundings of Sgr~A*. Subsequently, we put constraints on the charge of the Galactic centre black hole in Section~\ref{limits_charge}, including different processes that can induce charge and change its value, namely accretion of charged matter and the induction mechanism based on the black hole rotation in the magnetic field. In Section~\ref{section_observable_effects}, we focus on possible observational consequences of the charged SMBH, specifically the effect of charge on the black hole shadow size, the bremsstrahlung brightness profile, and the position of the innermost stable orbits of charged particles. We summarize the charge constraints in Section~\ref{section_summary_discussion}, where we discuss additional effects of the black-hole rotation and a potential non-electric origin of the charge. Finally, we conclude with Section~\ref{section_conclusions}.      

\section{Prospects for charging}
\label{section_prospects}

\begin{figure}
  \centering
  \includegraphics[width=0.5\textwidth]{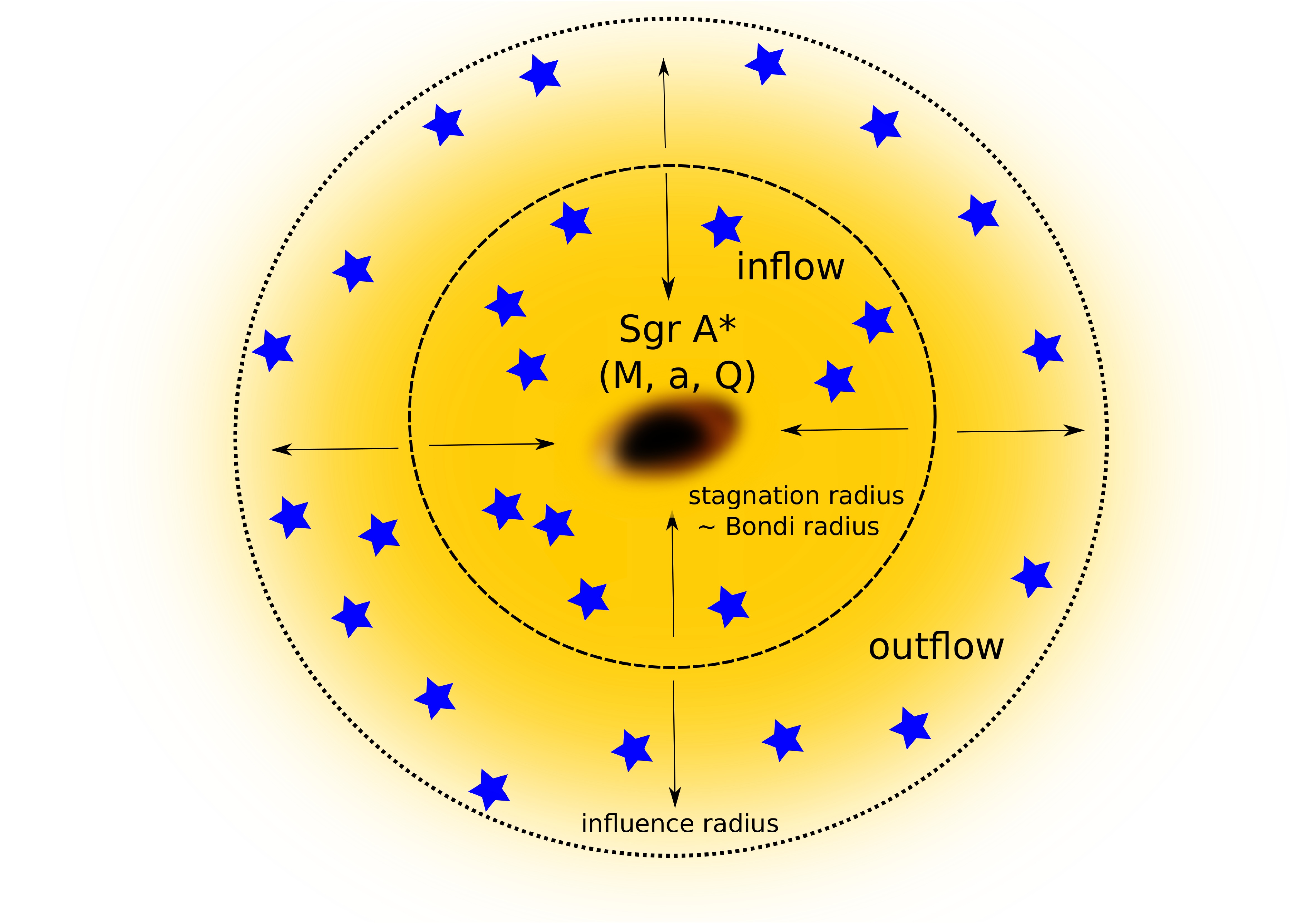}
  \caption{Illustration of the basic set-up at the Galactic centre: a supermassive black hole characterized by mass $M_{\bullet}$, spin $a_{\bullet}$, and an electric charge $Q_{\bullet}$ surrounded by a Nuclear star cluster and hot plasma emitting thermal bremsstrahlung. The dark area at the centre illustrates the shadow the black hole casts, which can be non-spherical due to the black hole rotation and the viewing angle. The inner circle denotes the stagnation radius, which is approximately equal to the Bondi radius, inside which the gas inflow towards the black hole takes place. The outer circle represents the sphere of gravitational influence of the supermassive black hole, inside which its potential prevails over the stellar cluster potential.}
  \label{fig_blackhole_plasma}
\end{figure}

Based on the analysis of surface brightness profiles in soft X-ray bands, there is an evidence for hot extended plasma, which surrounds the compact radio source Sgr~A* \citep{2010ApJ...716..504S,2013Sci...341..981W,2015A&A...581A..64R}. In addition, observations of polarized emission show that a relatively strong magnetic field is present in the Central Molecular Zone, which exhibits a highly ordered configuration \citep{2015llg..book..391M}. The large-scale ordered magnetic field as well as the ionized, extended gas surrounding Sgr~A* in the central region can be used to put constraints on the charge of the SMBH at the Galactic Centre, which has not been done properly before for any black hole candidate. The decreasing surface brightness profile is satisfactorily fitted by thermal bremsstrahlung \citep[see][for an analysis]{2015A&A...581A..64R}. The plasma is dynamically modelled in the framework of hot accretion flows, obtaining the temperature of $k_{\rm B} T_{\rm e}=1$, $2$, and $3.5\,{\rm keV}$ ($T_{\rm e}=(11.6-40.6)\times 10^6\,{\rm K}$) at the outer radius of the flow, using either the radiatively inefficient accretion flow (RIAF) model \citep{2013Sci...341..981W}, outflows of stars \citep{2010ApJ...716..504S}, or the classical Bondi accetion flow \citep{2015A&A...581A..64R}, respectively. Although several models are consistent with the observed surface brightness of plasma, both the RIAF and the Bondi accretion, which occur within the Nuclear Star Cluster, are expected to have a stagnation radius $R_{\rm stag}$ which divides the matter flowing in towards the SMBH and the outflowing gas \citep{2018MNRAS.479.4778Y}. In Fig.~\ref{fig_blackhole_plasma}, we illustrate the basic set-up, including the inflow, outflow region, and the stagnation radius. 

\subsection{Magnetic field properties}

There is an observational evidence for the highly ordered structure of the magnetic field in the central regions of the Galaxy \citep{2015llg..book..391M}. Two configurations were inferred from the observations of the polarized emission: a toroidal magnetic field associated with denser molecular clouds that is parallel with the Galactic plane and a poloidal field in the diluted intercloud region approximately perpendicular to the Galactic plane, which is also manifested by non-thermal radio filaments. The poloidal field in the intercloud region has magnitudes of $\sim 10\,{\rm \mu G}$ (close to the equipartition value with cosmic rays), and it reaches $\sim 1\,{\rm mG}$ in thin non-thermal filaments. The magnetic field in dense clouds has a toroidal geometry and it reaches the value of $\sim 1\,{\rm m G}$ \citep{2009A&A...505.1183F}.

Closer to Sgr~A*, \citet{2013Natur.501..391E} inferred the lower limit of magnetic field strength along the line of sight, $B \gtrsim 8\,{\rm mG}$, based on the Faraday rotation of the polarized emission of magnetar PSR J1745-2900, which is located at the deprojected distance of $r\gtrsim 0.12\,{\rm pc}$. In addition, they confirmed an ordered configuration of the magnetic field threading the hot plasma. Inside the stagnation radius, hot and magnetized plasma descends towards Sgr~A* and in this direction, an increase in the plasma density as well as in the magnetic field intensity is necessary. To explain the synchrotron emission of flares on the event-horizon scales, a magnetic field of the order of $\sim 10$--$100\,{\rm G}$ is required \citep{2000A&A...362..113F,2009ApJ...706..497M,2010ApJ...717.1092D,2012A&A...537A..52E}.  A simple scaling $B\propto r^{-1}$ is generally consistent with the increase from the Bondi-radius scales up to the event horizon.

\subsection{Plasma properties}

The inflow of plasma effectively takes place inside the Bondi radius, which gives the range of influence of the SMBH on the hot plasma,

\begin{equation}
  R_{\rm B}\approx 0.125 \left(\frac{M_{\bullet}}{4\times 10^6\,M_{\odot}}\right) \left(\frac{T_{\rm e}}{10^7\,{\rm K}} \right)^{-1}\left(\frac{\mu_{\rm HII}}{0.5}\right)\,{\rm pc}\,,
  \label{eq_Bondi_radius}
\end{equation}
where we assumed a fully ionized hydrogen plasma with the mean molecular weight of $\mu_{\rm HII}=0.5$ \citep{1978afcp.book.....L}. This assumption is supported by the observation of hot, ionized gas in the central arcsecond \citep{2015A&A...581A..64R}. In addition, at the inferred temperature of several keV and the number density of the order of $10\,{\rm cm^{-3}}$ at the Bondi radius \citep{2003ApJ...591..891B}, the ionization fraction of hydrogen atoms is basically unity according to the Saha equation, $(1-\chi)/\chi^2 \simeq 4.14 \times 10^{-16} n_{\rm tot} T_{\rm g}^{-3/2} \exp{(1.58\times 10^3\,{\rm K}/T_{\rm g})}$, where $\chi\equiv n_{\rm i}/n_{\rm tot}$ is the ionization fraction of the gas with the total gas number density of $n_{\rm tot}$ and temperature of $T_{\rm g}$. However, during the past high-luminosity states of Sgr~A* thermal instability could have operated in the inner parsec, creating the multi-phase environment where hot and cold phases could coexist \citep{2014MNRAS.445.4385R}. In addition, observations at millimeter wavelengths show the presence of both ionized and neutral/molecular medium in this region \citep[the denser and colder region is referred to as the minispiral,][]{2017A&A...603A..68M}. In the following, we will focus on the hot ionized phase, which is expected to dominate inside the Bondi radius. 

Plasma in the Galactic centre region is so hot inside the Bondi radius that it may be considered weakly coupled. This is easily shown by the coupling ratio $R_{\rm c}$ of the mean potential energy of particles and their kinetic energy,

\begin{equation}
 R_{\rm c}=\frac{E_{\rm p}}{E_{\rm k}}\sim \frac{e^2(L_{\rm i} 4\pi \epsilon_0)^{-1}}{k_{\rm B}T_{\rm e}}=  \frac{e^2n_{\rm p}^{1/3}(4\pi \epsilon_0)^{-1}}{k_{\rm B}T_{\rm e}}\,,
\end{equation}
where $L_{\rm i}$ is the mean interparticle distance, $L_{\rm i}=n_{\rm p}^{-1/3}$, where $n_{\rm p}$ is the particle density. For the typical (electron) particle density at the Bondi radius $n_{\rm p}\approx n_{\rm e} \approx 10\,{\rm cm^{-3}}$ and the electron temperature of $k_{\rm B}T_{\rm e}\sim 1\,{\rm keV}$ \citep{2003ApJ...591..891B,2013Sci...341..981W} as inferred from Chandra observations, we get $R_{\rm c}\approx 3 \times 10^{-10}$, i.e. the Galactic centre plasma is very weakly-coupled.

The Bondi radius of the accretion flow in the Galactic centre is thus well inside the sphere of influence of the SMBH, which represents the length-scale on which the potential of the SMBH prevails over the stellar cluster potential. For the Galactic centre SMBH and the averaged one-dimensional stellar velocity dispersion of $\sigma_{\star} \approx 100\,{\rm km\,s^{-1}}$, we get \citep[see e.g.][]{2013degn.book.....M,2015MNRAS.453..775G},
\begin{equation}
  R_{\rm inf}\simeq GM_{\bullet}/\sigma_{\star}^2=1.7 \left(\frac{M_{\bullet}}{4\times 10^6\,M_{\odot}} \right) \left(\frac{\sigma_{\star}}{100\,{\rm km\,s^{-1}}} \right)^{-2}\,{\rm pc}\,.
  \label{eq_influence_radius}
\end{equation}



When considering a one-dimensional steady-state inflow-outflow structure of the gas in the vicinity of a galactic nucleus, a characteristic feature is the existence of the stagnation radius $R_{\rm stag}$, where the radial velocity passes through zero \citep{2015MNRAS.453..775G}. Stellar winds inside the stagnation radius flow towards the black hole and a fraction of the matter is accreted, while the matter outside it forms an outflow, which is illustrated in Fig.~\ref{fig_blackhole_plasma}. For the case when the heating rate due to fast outflows $v_{\rm w}$ is larger than the stellar velocity dispersion $\sigma_{\star}$, $v_{\rm w} \gg \sigma_{\star}$, the stagnation radius can be approximately expressed as \citep{2015MNRAS.453..775G},

\begin{align}
  R_{\rm stag} & \approx \left(\frac{13+8\Gamma}{4+2\Gamma}-\frac{3\nu}{2+\Gamma}\right)\frac{GM_{\bullet}}{\nu v_{\rm w}^2}\,\notag\\
            & \approx
 \begin{cases}
   0.30\,\left(\frac{M_{\bullet}}{4\times 10^6\,M_{\odot}}\right)\left(\frac{v_{\rm w}}{500\,{\rm km\,s^{-1}}}\right)^{-2}\,{\rm pc} &,  \text{core\, ($\Gamma=0.1$)}\,,\\
   0.16\,\left(\frac{M_{\bullet}}{4\times 10^6\,M_{\odot}}\right)\left(\frac{v_{\rm w}}{500\,{\rm km\,s^{-1}}}\right)^{-2}\,{\rm pc} &,  \text{cusp\, ($\Gamma=0.8$)}\,,
  \end{cases}
  \label{eq_stagnation_radius}
\end{align} 
where $\Gamma$ is the inner power-law slope of the stellar brightness profile, where we consider two limiting cases, the core profile with $\Gamma=0.1$ and the cusp profile with $\Gamma=0.8$. The quantity $\nu=-\mathrm{d}\rho/\mathrm{d}r|_{\rm R_{stag}}$ is the gas density power-law slope at $R_{\rm Stag}$, which according to the numerical analysis of \citet{2015MNRAS.453..775G} is $\nu \approx 1/6[(4\Gamma+3)]$. According to the estimates in Eq.~\eqref{eq_stagnation_radius}, the stagnation radius is expected to be nearly coincident with the Bondi radius with an offset given by the factor \citep{2015MNRAS.453..775G}

\begin{equation}
  \frac{R_{\rm Stag}}{R_{\rm B}} \approx \frac{13+8\Gamma}{(2+\Gamma)(3+4\Gamma)}\,,
  \label{eq_stag_Bondi}
\end{equation}
which is of the order of unity.

In the further analysis and estimates, we consider the dynamical model of \citet{2015A&A...581A..64R}, who fitted the thermal bremsstrahlung emission of the hot plasma with the classical Bondi solution. They found that the Bondi solution can reproduce well the surface brightness profile up to the outer radius of $r_{\rm out} \sim 3'' \approx R_{\rm B}$ (where $1''\approx 0.04\,{\rm pc}$), which is consistent with the Bondi radius expressed in Eq.~\eqref{eq_Bondi_radius}.

Considering the surface brightness profile of the hot flow inferred from 134 ks Chandra ACIS-I observations, the best fitted model of the spherical Bondi flow gives the asymptotic values of the electron density of $n_{\rm e}^{\rm out}=18.3 \pm 0.1\,{\rm cm^{-3}}$, the electron temperature of $T_{\rm e}^{\rm out}=3.5\pm 0.3\,{\rm keV}$, and the sound speed of $c_{\rm s}^{\rm out}=7.4\times 10^7\,{\rm cm\,s^{-1}}$. The steady spherical Bondi solution gives power-law profiles for the electron density and the electron temperature inside the Bondi radius,

\begin{align}
  n_{\rm e}&\approx n_{\rm e,0} \left(\frac{r}{r_0}\right)^{-3/2}\,,\notag\\
  T_{\rm e}&\approx T_{\rm e,0} \left(\frac{r}{r_0}\right)^{-1}\,,
  \label{eqs_powerlaws_bondi}
\end{align}
where $n_{\rm e,0}=70\,{\rm cm^{-3}}$ and $T_{\rm e,0}=9\,{\rm keV}$ at $r_0=0.4''$ \citep{2015A&A...581A..64R}. The temperature has a virial profile for the adiabatic index of $\gamma_{\rm ad}=5/3$.

The importance of collisional processes between plasma constituents -- mainly protons and electrons -- can be evaluated by comparing the collisional timescales of electron-electron and electron-proton interactions with the typical dynamical timescale (free-fall timescale) and the viscous timescale. The electron-electron collisional frequency approximately is $\nu_{\rm ee} \simeq 3.75 n_{\rm e} T_{\rm e}^{-3/2} \log{\Lambda_{\rm ee}}\,{\rm Hz}$, where $\log{\Lambda_{\rm ee}}$ is the Coulomb logarithm, while the frequency of electrons colliding with protons may be estimated as $\nu_{\rm ep} \simeq 5.26 n_{\rm i} T_{\rm e}^{-3/2} \log{\Lambda_{\rm ei}}\,{\rm Hz}$. Assuming that the electron and proton number densities are approximately the same, $n_{\rm i} \approx n_{\rm e}$, the corresponding collisional timescales of electrons with themselves and protons are approximately constant with radius, given the Bondi profiles in Eqs.~\eqref{eqs_powerlaws_bondi}. The electron-electron collisional timescale is $\tau_{\rm ee}\approx (\nu_{\rm ee})^{-1} \approx 13\,{\rm yr}$ for $\log{\Lambda_{\rm ee}}=10$. In a similar way, the electron-proton collisional timescale is  $\tau_{\rm ep}\approx 9\,{\rm yr}$. The free-fall (dynamical) timescale evaluated for the initial infall distance of $r_0=1000 r_{\rm S}$ is,

\begin{equation}
  t_{\rm dyn}\approx t_{\rm ff} \simeq 0.062\,\left(\frac{r_0}{1000\,r_{\rm S}}\right)^{3/2}\left(\frac{M_{\bullet}}{4\times 10^6\,M_{\odot}}\right)^{-1/2}\,.
  \label{eq_dynamical_timescale}
\end{equation}
The viscous timescale $t_{\rm vis}$ may be in general expressed as \citep{2016ASSL..440....1L},
\begin{equation}
  t_{\rm vis} \approx \alpha^{-1} (H/r_0)^{-2} t_{\rm dyn}\,
  \label{eq_viscous_timescale}
\end{equation}
where the ratio of the thickness of the accretion flow to the radial length-scale is expected to be of the order of unity since the accretion flow around Sgr~A* is generally considered to be optically thin and geometrically thick as for hot accretion flows in general \citep{2014ARA&A..52..529Y}. When the viscosity parameter $\alpha$ is of the order of $0.1$ \citep{2007MNRAS.376.1740K,2016ASSL..440....1L}, the viscous timescale is $t_{\rm vis} \approx 10t_{\rm dyn}$. 

\begin{figure}
  \centering
  \includegraphics[width=0.5\textwidth]{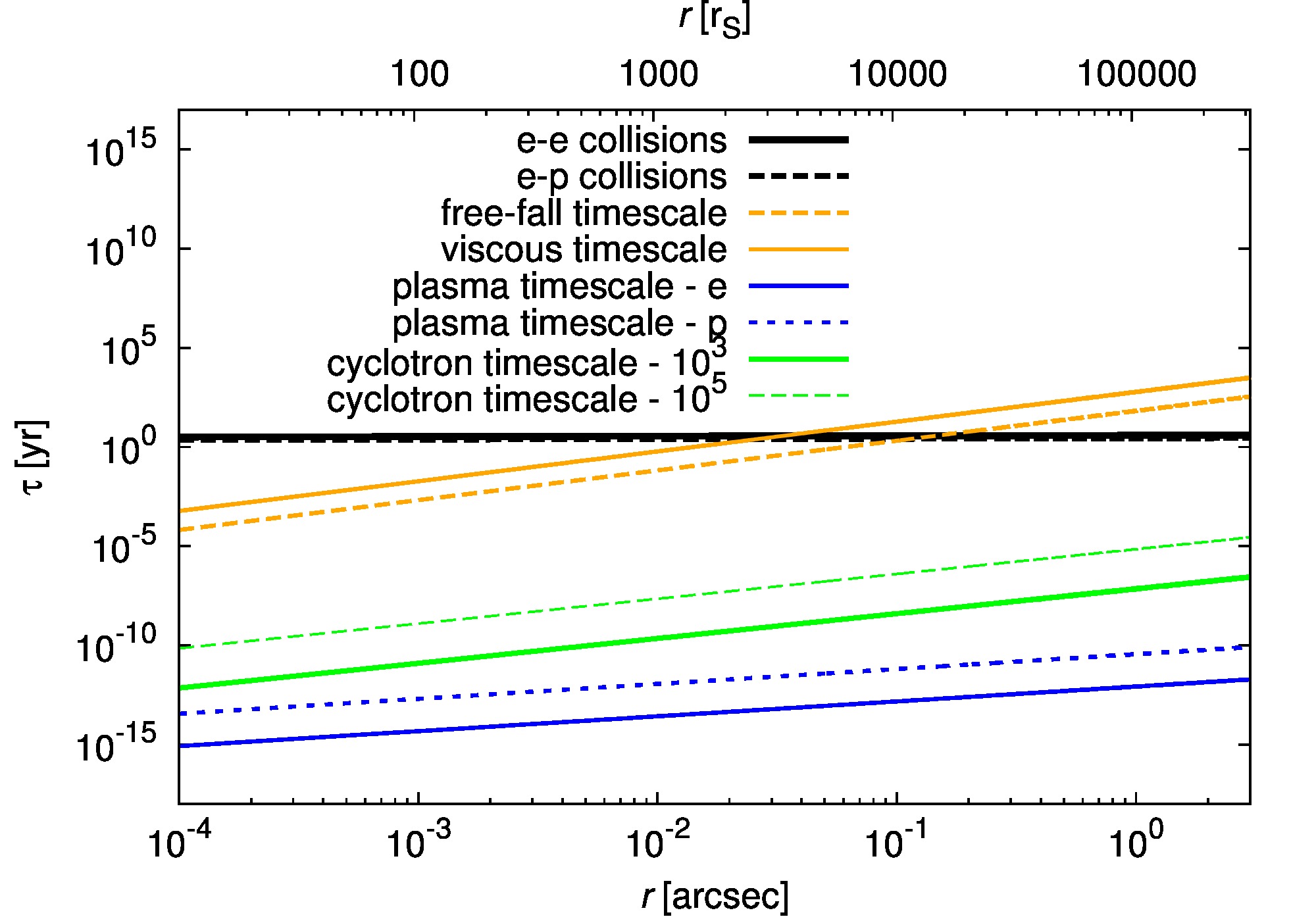}
  \caption{Radial profiles of electron-electron and electron-proton collisional timescales estimated based on density and temperature radial profiles expressed by Eqs.~\ref{eqs_powerlaws_bondi}. Inside the innermost 1000 Schwarzschild radii, the electron-electron and electron-proton collisions occur on longer timescales than the dynamical and accretion processes governed by Sgr~A*. On the other hand, the density and velocity plasma oscillations take place on significantly shorter scales throughout the GC region. The same applies to the range of cyclotron timescales.}
  \label{fig_collisional_timescales}
\end{figure}

When compared to the free-fall and viscous timescales, see Fig.~\ref{fig_collisional_timescales}, the electron-electron and electron-proton interactions take place on longer timescales than the dynamical and accretion processes inside the inner 1000 Schwarzschild radii. This implies that particle collisions are irrelevant for the dynamical processes in the immediate vicinity of Sgr~A*.

On the other hand, the ordered plasma oscillations with the characteristic plasma frequency $\nu_{\rm p}$ are relevant on all spatial scales. The presence of plasma close to the SMBH at the Galactic centre means that the immediate vicinity of Sgr~A* cannot be observed at frequencies smaller than the plasma frequency, $\nu < \nu_{\rm p}$, because of charge oscillations in the plasma. The Bondi-flow model of \mbox{\citet{2015A&A...581A..64R}} predicts the number densities of electrons at the scale of $r=10\,r_{\rm S}$ to be $n_{\rm e}\approx 10^7\,{\rm cm^{-3}}$ and that yields the plasma frequency of,

\begin{equation}
  \nu_{\rm p}= 28.4\,\left(\frac{n_{\rm e}}{10^7\,{\rm cm^{-3}}} \right)^{1/2}\,{\rm MHz}\,,
  \label{eq_plasma_freq}
\end{equation}
which would effectively block electromagnetic radiation with wavelengths longer than $\lambda_{\rm p}\simeq 11\,{\rm m}$ from the innermost region. By coincidence, the plasma frequency close to the Galactic centre expressed by Eq.~\eqref{eq_plasma_freq}, which depends on the electron density, is close to the cyclotron frequency for electrons gyrating in the magnetic field with the intensity of $B\sim 10\,{\rm G}$,
 \begin{equation}
    \nu_{\rm cyc}=\frac{B}{2\pi\gamma_{\rm L}}\frac{e}{m_{\rm e}}\simeq 28 \left(\frac{B}{10\,{\rm G}}\right)\gamma_{\rm L}^{-1}\,{\rm MHz}\,,
    \label{eq_cyclotron_freq}
 \end{equation}
 where $\gamma_{\rm L}$ is a Lorentz factor. The approximate profile of the cyclotron timescale can be evaluated using the assumption that the magnetic field pressure is a fraction of the gas pressure, $P_{\rm gas}=n_{\rm e}k_{\rm B}T_{\rm e}$. Then the magnetic field is $B=\sqrt{8 \pi P_{\rm gas}/\beta}$, where we take $\beta=100$ to reproduce the magnetic field strengths as determined based on the magnetar observations at larger distances and the flare observations on the ISCO scales \citep{2012A&A...537A..52E,2013Natur.501..391E}. The Lorentz factor is taken in the range $\gamma_{\rm L}=10^3-10^5$ as inferred from the flare observations in the X-ray and infrared domains \citep{2012A&A...537A..52E}. Based on Fig.~\ref{fig_collisional_timescales}, both the plasma timescale and the cyclotron timescales for electrons are shorter than the dynamical, viscous, and collisional timescales in the whole considered region.

Although particle collisions can be neglected for dynamical processes in the vicinity of Sgr~A*, this does not apply to radiative processes in the same region, namely for the detected thermal X-ray bremsstrahlung \citep{2003ApJ...591..891B,2018MNRAS.474.3787M}. By definition, the bremsstrahlung requires the Coulomb interaction of particles, with the dominant contribution of unlike particles, electrons and protons. This is not in contradiction with the low estimated rate of collisions in the central region, since the interactions over the whole region contribute to the observed surface brightness.

The average electron and proton (ion) density can be estimated from the emissivity of thermal bremsstrahlung and the quiescent X-ray luminosity of Sgr A*. In principle, the thermal bremsstrahlung can experience losses due to Thomson scattering. The optical depth along the line of sight may be estimated as $\tau_{\rm T}= \int_0^{R_{\rm B}} \sigma_{\rm T} n_{\rm e}(l) \mathrm{d}l$, where $l$ is the line-of-sight coordinate. Given the cross-section of Thomson scattering, $\sigma_{\rm T}=6.65\times 10^{-25}\,{\rm cm^{2}}$, the electron density profile given by Eq.~\eqref{eqs_powerlaws_bondi}, and the typical length-scale given by the Bondi radius, $R_{\rm B}$, the optical depth is given by
\begin{equation}
  \tau_{\rm Brems}=\int_0^{R_{\rm B}} \sigma_{\rm T} n_{\rm e}(l) \mathrm{d}l=\frac{\sigma_{\rm T} n_{\rm e,0} r_0^{3/2}}{R_{\rm B}^{1/2}}\approx 2\times 10^{-7}\,,
  \label{eq_depth_thomson}
\end{equation}
and hence the losses due to Thomson scattering are negligible. Given the average quiescent X-ray luminosity of Sgr~A* in the range of $2$--$10\,{\rm keV}$, $L_{2-10}=2\times 10^{33}\,{\rm erg\,s^{-1}}$ \mbox{\citep{2017IAUS..322....1H}}, one can estimate the electron density from the thermal bremsstrahlung luminosity,

\begin{align}
  L_{\rm brems}\approx & 6.8\times 10^{-38} Z^2 n_{\rm i} n_{\rm e} T_{\rm g}^{-1/2} V(R_{\rm B}) g_{\rm ff}(\nu,T_{\rm g}) \times \,\notag \\ 
                       &\times \int_{\nu_1}^{\nu_2} \exp{(-h\nu/kT_{\rm g})} \mathrm{d}\nu\,{\rm erg\,s^{-1}}\,,
  \label{eq_brems_lum}
\end{align}
where $n_{\rm i}$ is the ion number density, $Z$ is the proton number of participating ions, $T_{\rm g}$ is the temperature of the gas, $V(R_{\rm B})\approx 4/3\pi R_{\rm B}^3$ is the volume inside the Bondi radius, $g_{\rm{ff}}(\nu, T_{\rm g})$ is a Gaunt factor used for the quantum corrections to classical formulas, which we set to $g_{\rm{ff}}=1.5$ in the given temperature range of $(10^7,10^8)\,{\rm K}$ and the frequency range of $(2,10)\,{\rm keV}=(\nu_1,\nu_2)=(0.5,2.4)\times 10^{18}\,{\rm Hz}$.
The integral in Eq.~\eqref{eq_brems_lum} can be approximated as $ \int_{\nu_1}^{\nu_2} \exp{(-h\nu/kT_{\rm g})} \mathrm{d}\nu \sim 0.98 \times 10^{10}\,T_{\rm g}$. For fully ionized hydrogen and helium plasma, the term $Z^2 n_{\rm i} n_{\rm e}$ becomes $1.55n_{\rm e}^2$. Putting the numerical factors all together into Eq.~\eqref{eq_brems_lum} yields,

\begin{equation}
 L_{\rm brems}\approx 3.725\times 10^{30} \overline{n}_{{\rm e}}^2 \left(\frac{T_{\rm g}}{10^8\,{\rm K}}\right)^{1/2}\left(\frac{R_{\rm B}}{0.125\,{\rm pc}}\right)^{3}\,{\rm erg\,s^{-1}}\,.
 \label{eq_brems_lum2}
\end{equation}
Given the measured X-ray luminosity of $L_{2-10}\approx 2\times 10^{33}\,{\rm erg\,s^{-1}}$, the mean electron density using Eq.~\eqref{eq_brems_lum2} is $\overline{n}_{\rm e} \approx 23\,{\rm cm^{-3}}$, which is very close to the asymptotic value of $n_{\rm e}^{\rm out}=18.3 \pm 0.1\,{\rm cm^{-3}}$ derived by \citet{2015A&A...581A..64R} from the Bondi solution. 

\subsection{Classical estimates of charging}
\label{sec_classical_estimates}

The fundamental mechanism, which can lead to charging of the black hole, are different thermal speeds for electrons and protons in the fully ionized plasma, $v_{\rm th,e}=(k_{\rm B} T_{\rm e}/m_{\rm e})^{1/2}$ and $v_{\rm th,p}=(k_{\rm B} T_{\rm p}/m_{\rm p})^{1/2}$, following from the fact that the Galactic centre plasma is collisionless. Given a considerable mass difference between electrons and protons, $m_{\rm p}=1837 m_{\rm e}$, it is expected that the ratio of thermal speeds is,

\begin{equation}
  \frac{v_{\rm th,e}}{v_{\rm th,p}}=\left(\frac{T_{\rm e}}{T_{\rm p}}\frac{m_{\rm p}}{m_{\rm e}}\right)^{1/2} \approx 43\,,
  \label{eq_sound_speed_ratio}
\end{equation}
under the assumption that $T_{\rm e}\approx T_{\rm p}$. This leads to the ratio of the corresponding Bondi radii for protons and electrons,

\begin{equation}
  \frac{R_{\rm B,p}}{R_{\rm B,e}}=\frac{T_{\rm e}}{T_{\rm p}}\frac{m_{\rm p}}{m_{\rm e}}\,.
  \label{eq_ratio_bondi}
\end{equation}

Next, we can estimate the total charge by integrating across the corresponding Bondi radius. In the spherical symmetry, the total charge inside the Bondi radius can be calculated as $|Q|=\int_0^{R_{\rm B}} \rho_{Q} 4\pi r^2 \mathrm{d}r$, where $\rho_{Q}$ is the charge density. Under the assumption of a power-law density profile for both electrons and protons, $n_{\rm e,p}=n_0 (r/r_0)^{-\gamma_{\rm n}}$ ($\gamma_{\rm n}=3/2$ for the spherical Bondi flow), the ratio of positive and negative charge in the range of influence of the SMBH can be calculated as follows,

\begin{align}
  \left|\frac{Q_{+}}{Q_{-}}\right|&=\frac{\int_0^{R_{\rm B,p}}e n_{\rm p}4\pi r^2\mathrm{d}r}{\int_0^{R_{\rm B,e}}e n_{\rm e}4\pi r^2\mathrm{d}r}\,,\\ \notag
                       &=\left(\frac{R_{\rm B,p}}{R_{\rm R_{B,e}}}\right)^{3-\gamma_{\rm n}}\,,\\ \notag
                       &=\left(\frac{T_{\rm e}}{T_{\rm p}}\frac{m_{\rm p}}{m_{\rm e}}\right)^{3-\gamma_{\rm n}}\approx 8\times 10^4\,.
\end{align} 
where the last estimate is valid for $T_{\rm e} \approx T_{\rm p}$, which, however, does not have to be quite valid in the hot accretion flows that surround quiescent black holes, such as Sgr~A* or M87, where a two-temperature flow is expected to exist \citep{2014ARA&A..52..529Y}. At large distances from the black hole, close to the stagnation radius, the electron and proton temperatures are expected to be almost the same. Closer to the black hole, in the free-fall regime of the Bondi flow, the electron and the proton (ion) temperatures differ, the proton temperature is larger than the electron temperature by about a factor of $\simeq 1-5$, $T_{\rm e}/T_{\rm p}\geq 1/5$ \citep{2009ApJ...706..497M,2010ApJ...717.1092D}, which gives the lower limit to the excess of the positive charge in the range of influence of the SMBH, $\left|\frac{Q_{+}}{Q_{-}}\right|\gtrsim 7000$.

A similar analysis as above was discussed and performed for stationary plasma atmospheres of stars \citep[see][and references therein]{2001A&A...372..913N} and in general, for gravitationally bound systems of plasma \citep{1978ApJ...220..743B}. In the hot atmosphere, where the plasma may be considered collisionless, lighter electrons tend to separate from heavier protons. The separation is stopped by an induced electrostatic field $\psi=(1/4 \pi \epsilon_0)Q_{\rm eq}/r$. In the gravitational field of an approximately spherical mass of $M_{\bullet}$, $\phi=GM_{\bullet}/r$, the potential energy of protons can be expressed as $E_{\rm p}=-m_{\rm p}\phi +e\psi$, and for electrons with negative charge in a similar way, $E_{\rm e}=-m_{\rm e}\phi -e\psi$. Given the assumption of the stationary equilibrium plasma densities, the number densities of protons and electrons are proportional to $\exp{(-E_{\rm p}/k_{\rm B}T_{\rm p})}$ and $\exp{(-E_{\rm e}/k_{\rm B}T_{\rm p})}$, respectively. Given the quasineutrality of plasma around stars and Sgr~A*, the difference between the densities of protons and electrons is expected to be small, which implies $E_{\rm p} \approx E_{\rm e}$. The induced equilibrium charge of the central body surrounded by plasma then is,

\begin{align}
  Q_{\rm eq}  &= \frac{2\pi \epsilon_0 G(m_{\rm p} - m_{\rm e})}{e}M_{\bullet}\,\notag\\                 &\approx 3.1 \times 10^8 \left(\frac{M_{\bullet}}{4\times 10^6\,M_{\odot}} \right)\,C\,.
  \label{eq_charge_blackhole}
\end{align}
Eq.~\eqref{eq_charge_blackhole} was derived under the assumption of spherical stationary plasma around a point mass, which is certainly not met in the environment of dynamical plasma around Sgr~A*. The real charge of Sgr~A* will therefore deviate from the stationary value of $Q_{\rm eq}$. The mechanism of charging will, however, still tend to operate and a rough approximation of charge expressed by Eq.~\eqref{eq_charge_blackhole} is still more precise than the assumption of a neutral black hole. The equilibrium value $Q_{\rm eq}$ also expresses the upper limit of an electric charge associated with Sgr~A*, for which the stationary number densities of protons and electrons in plasma remain approximately constant. For charges $Q \gg Q_{\rm eq}$, a significant difference in the number densities is expected that would decrease with the distance, unless the charge of the black hole would be Debye-shielded, as we will discuss later in this paper.

Given the simple calculations above, it is expected that the black hole at the Galactic centre can acquire a small positive charge, given the fact that more positive charge is in the range of its gravitational influence than negative charge. In the following, we will look at more realistic scenarios of how the black hole charge could be induced in the accretion flow, given the fact that black holes are expected to have a non-zero spin and should be immersed in a magnetic field.

\section{Limits on the black hole charge}
\label{limits_charge}

\subsection{Maximum theoretical values of the charge of Sgr~A*}

The uppermost limit on the charge of Sgr~A* may be derived using the the space-time of the black hole that is characterized by its mass $M_{\bullet}$, electric charge $Q_{\bullet}$, and rotation parameter $a_{\bullet}$. In the most general case, the Kerr-Newman (KN) solution \citep{1965JMP.....6..918N} fully describes such a black hole in vacuum. The KN metric can be expressed in Boyer-Lindquist coordinates in the following way \citep{1973grav.book.....M},
\begin{align}
  \mathrm{d}s^2 = &-\left(\frac{\mathrm{d}r^2}{\Delta}+\mathrm{d}\theta^2\right)\rho^2+\left(c\mathrm{d}t-a_{\bullet}\sin^2{\theta}\mathrm{d}\Phi\right)^2\frac{\Delta}{\rho^2}-\nonumber\\
  &-[(r^2+a_{\bullet}^2)\mathrm{d}\Phi-a_{\bullet}c\mathrm{d}t]^2\frac{\sin^2{\theta}}{\rho^2}\,,
  \label{eq_kerr_newman}
\end{align}
where $\rho^2=r^2+a_{\bullet}^2\cos^2{\theta}$ and $\Delta=r^2-r_{\rm S}r+a_{\bullet}^2+r_{\rm Q}^2$. The length-scales $r_{\rm S}$ and $r_{\rm Q}^2$ correspond to the Schwarzschild radius $r_{\rm S}=2GM_{\bullet}/c^2=1.2 \times 10^{12} (M_{\bullet}/4\times 10^6\,M_{\odot})\,{\rm cm}$ and $r_{\rm Q}^2=GQ_{\bullet}^2/(4\pi\epsilon_0c^4)$, respectively. The position of the event horizons is obtained with the condition $\Delta=0$, which leads to the quadratic equation, $r^2-r_{\rm S}r+a_{\bullet}^2+r_{\rm Q}^2=0$, with two possible horizons $r_{1,2}=1/2(r_{\rm S}\pm\sqrt{r_{\rm S}^2-4(a_{\bullet}^2+r_{\rm Q}^2)})$ for $r_{\rm S}>2\sqrt{a_{\bullet}^2+r_{\rm Q}^2}$. For $r_{\rm S}<2\sqrt{a_{\bullet}^2+r_{\rm Q}^2}$, no horizons exist, so the object is a naked singularity. The condition $r_{\rm S}=2\sqrt{a_{\bullet}^2+r_{\rm Q}^2}$ leads to a single event horizon located at $r=1/2r_{\rm S}$, which represents an extremal black hole, and it also gives an upper limit for an electric charge of the SMBH at the Galactic centre,

\begin{equation}
  Q_{\rm max}^{{\rm rot}}=2c^2\sqrt{\frac{\pi\epsilon_0}{G}\left(\frac{G^2 M_{\bullet}^2}{c^4}-a_{\bullet}^2 \right)}\,
  \label{eq_max_gen}
\end{equation}
which can be rewritten using a dimensionless parameter $a_{\bullet}=\tilde{a}_{\bullet}GM_{\bullet}/c^2$ into the form,
\begin{equation}
  Q_{\rm max}^{{\rm rot}}=2M_{\bullet}\sqrt{\pi \epsilon_0 G (1-\tilde{a}_{\bullet}^2)}\,.
  \label{eq_max_gen_nodim}
\end{equation}

Relation~\eqref{eq_max_gen} represents a theoretical limit for the maximum charge of a rotating black hole. Above this limit, the massive object at the Galactic centre would be not a black hole anymore, but a naked singularity, which can be ruled out based on observational and causal arguments \citep{2017FoPh...47..553E}. In addition, a direct transition between a non-extremal black hole and an extremal one due to the accretion of charged matter (test particles or shells) is not possible as was shown in \citet[][see also references therein]{1998PhRvD..57.5284W}. 

For a non-rotating black hole $(a_{\bullet}=0)$, the maximum charge is proportional to the black hole mass. Evaluating for Sgr~A* gives,

\begin{equation}
 Q_{\rm max}^{{\rm norot}}=2\sqrt{\pi \epsilon_0 G}M_{\bullet}=6.86 \times 10^{26}\, \left(\frac{M_{\bullet}}{4\times 10^6\,M_{\odot}} \right)\,C\,.
 \label{eq_max_charge}
\end{equation}

\begin{figure*}
  \centering
  \begin{tabular}{cc}
  \includegraphics[width=0.5\textwidth]{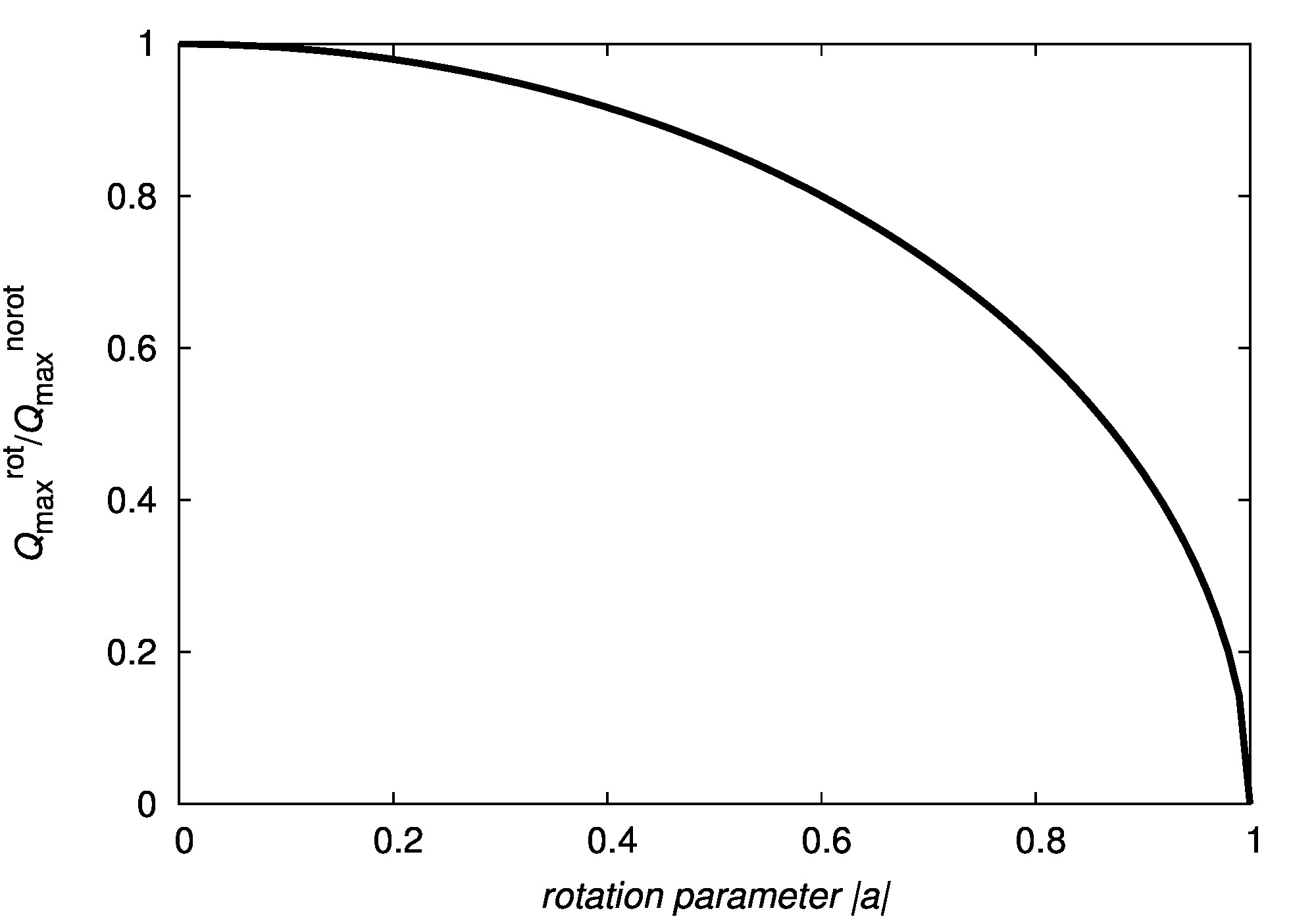} & \includegraphics[width=0.5\textwidth]{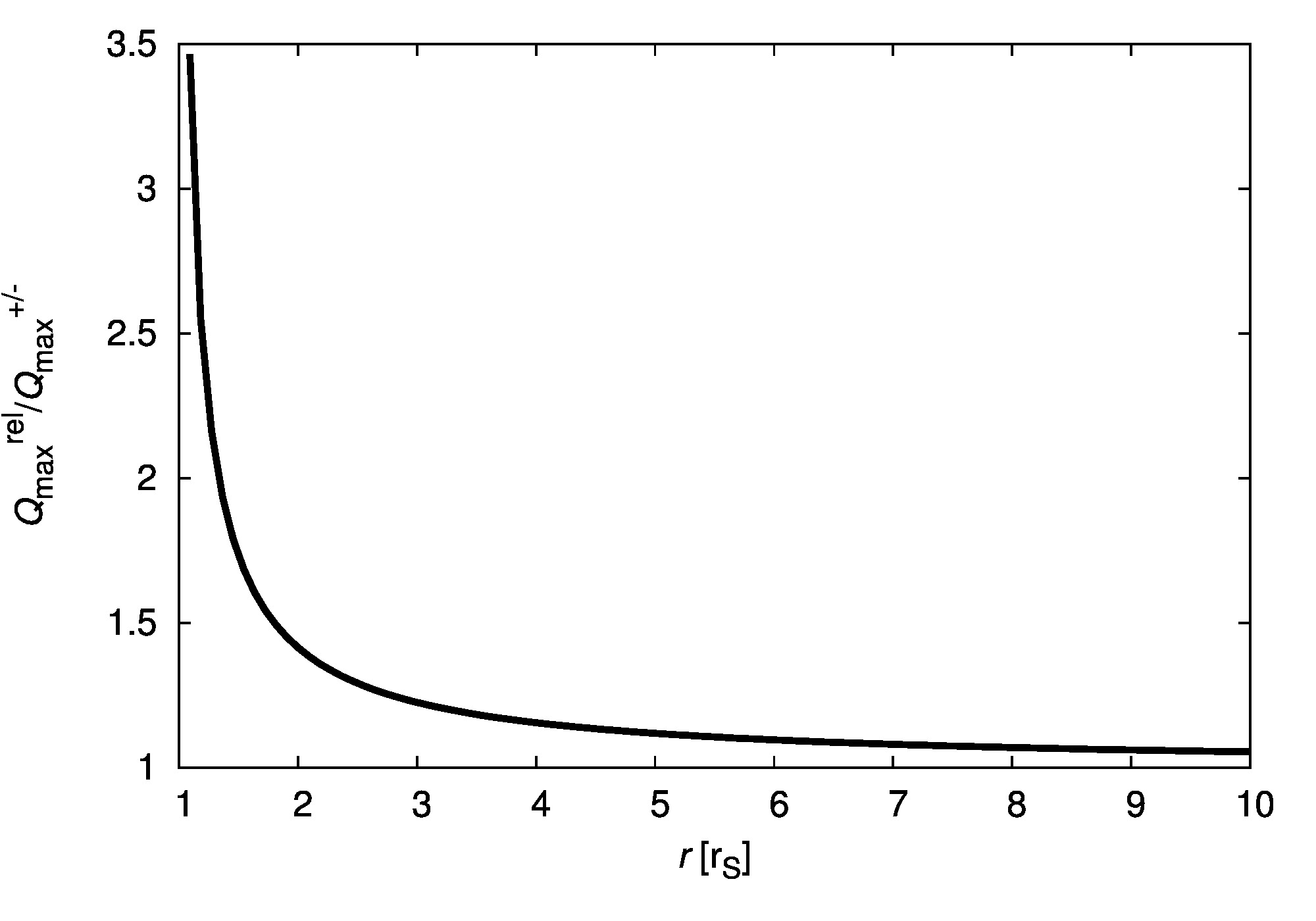}
  \end{tabular}
  \caption{\textbf{Left:} Effect of the rotation of the black hole on its maximum electric charge. \textbf{Right:} The dependence of the relativistic correction to both the maximum positive and negative charge of the supermassive black hole. In the relativistic regime, the electrostatic force increases towards the event horizon, which becomes apparent inside the innermost stable orbit.}
  \label{fig_rel_correction}
\end{figure*}

In Fig.~\ref{fig_rel_correction}, we plot the effect of the rotation of the black hole on its maximum electric charge. Close to the maximum rotation, the maximum charge is close to zero. 

In the following, we discuss limits on the electric charge of the Galactic centre black hole based on induced electric field of a rotating black hole immersed in the circumnuclear magnetic field.

\subsection{Charge induced by rotating SMBH}
\label{subsec_rotating_SMBH}

There are indications that a considerable magnetic field must be present in the vicinity of SMBH at the Galactic centre \citep{2012A&A...537A..52E,2013Natur.501..391E,2015A&A...581A..64R}, which suggests the value of $10\,$G in the vicinity of the even horizon. The exact solution for the electromagnetic fields in the vicinity of Sgr~A* is far from being properly defined, however it is natural to assume that the magnetic field shares the symmetries of the background spacetime metric, such as the  axial symmetry and stationarity. The central SMBH is assumed to be a standard Kerr black hole whose gravitational potential dominates the system. Moreover, the small density of the plasma around Sgr~A* implies that the magnetospere of SMBH can be described within the test field approximation.
 This assumption implies that the four-vector potential can be written in the form $A^{\alpha}=k_1\xi^{\alpha}_{(t)}+k_2\xi^{\alpha}_{(\phi)}$, where $k_1$ and $k_2$ are the arbitrary parameters and $\xi^{\alpha}_{(t)}$ and $\xi^{\alpha}_{(\phi)}$ are a unit timelike and spacelike Killing vectors, respectively, which are related to the symmetries of the black hole spacetime.
 For estimation purposes, one can assume that the magnetic field is homogeneous and aligned along the axis of rotation of black hole with the strength $B$. Then, the solution of Maxwell equations for corresponding four-vector potential in rotating black hole spacetime can be written in the following form \citep{1974PhRvD..10.1680W} 
\beq 
A_t = \frac{B}{2} \left(g_{t\phi} + 2 a g_{tt}\right), \quad A_{\phi} =  \frac{B}{2} \left(g_{\phi\phi} + 2 a g_{t\phi}\right).
\label{VecPotTP}
\eeq
%
%
The rotation of a black hole gives the contribution to the Faraday induction which generates the electric potential $A_t$ and thus produces an induced electric field,  just in the same manner as if the field would be induced in magnetic field by a rotating ring. This is how the black hole obtains nonzero induced charge. The potential difference between the horizon of a black hole and infinity takes the form 
\beq
\Delta \phi = \phi_{\rm H} - \phi_{\infty} = \frac{Q - 2 a M B}{2 M}.
\eeq
The potential difference leads to the process of selective accretion, which implies that a rotating black hole in external homogeneous magnetic field can obtain maximum net electric charge $Q= 2 a M B$.

The statement of selective accretion by rotating black hole in the presence of magnetic field is quite general and independent of the exact shape of the field and the type of accreting charged matter, which can, however, put restrictions on the timescales of selective accretion due to charge separation in a plasma. In general, the energy of charged particle in stationary field and spacetime is $E=-P_{\mu} \xi^{\mu}_{(t)}$, where $P_{\mu} = m u_{\mu} - q A_{\mu}$ is generalized four-momentum. The difference of electrostatic energy of a particle at infinity and at the horizon is $E_H - E_{\infty} = q A_{t|r\rightarrow r_H} - q A_{t|r\rightarrow \infty} = \delta$.  When $\delta>0$, the black hole accretes the particles with $q$ charges, while when $\delta<0$, it is energetically more favourable to accrete the particles with $-q$ charge. In both cases the black hole will accrete a net charge until the difference $\delta$ is reduced to zero. The sign of induced charge depends on the relative orientation of magnetic field lines and black hole spin \citep{2016PhRvD..93h4012T}. If one assumes that the magnetic field is created by the dynamics of the surrounding conducting plasma which co-rotates with the black hole, then the sign of induced charge is more likely positive.
%
%

This analysis can also put a limit on the maximum induced charge of SMBH. If magnetic field is oriented along the rotation axis of a black hole, the black hole induces electric charge given by $Q_{\bullet \rm ind} = 2 a M_{\bullet} B_{\rm ext}$.  Given an upper boundary for the spin parameter  $a_{\bullet} \leq M_{\bullet}$ one can estimate the upper boundary for the induced charge as follows
\beq 
Q_{\bullet \rm ind}^{\rm max} = 2.32 \times 10^{15} \left( \frac{M_{\bullet}}{4 \times 10^6 M_{\odot}} \right)^2  \left( \frac{B_{\rm ext}}{10 \rm G} \right)  ~\rm C,
\eeq
where the external magnetic field is expressed in units of $10$ Gauss, which is the estimated magnetic field strength associated with the flaring activity of Sgr~A* \citep{2012A&A...537A..52E}.
Independently from the precise configuration of magnetic field in the Galactic center, the order of magnitude of estimated induced charge is about $10^{15 \pm 1}$C, if the assumptions of the axial symmetry and stationarity is preserved. In the case of Sgr~A*, there is a convincing evidence that the magnetic field in the Galactic center is indeed highly oriented and ordered \citep{2015llg..book..391M}, which supports our assumptions.

We would like to stress that the induced charge plays a role of a driving force in the processes of energy extraction from rotating black holes. Two leading processes by which the rotational energy can be extracted out are the Blandford-Znajek mechanism \citep{1977MNRAS.179..433B} and the magnetic version of Penrose process \citep{1969NCimR...1..252P,1985ApJ...290...12W}, both of which use the existence of negative energy states of particles with respect to observer at infinity. In both of these processes, the rotation of a black hole in magnetic field generates quadrupole electric field by twisting of magnetic field lines \citep{2018MNRAS.478L..89D}. An infall of oppositely charged particles (relative to the sign of induced charge) leads to the discharge of the induced field and therefore the extraction of rotational energy of a black hole. In Blandford-Znajek mechanisms the presence of induced field completes current circuit for in-falling oppositely charged negative energy flux. This field is responsible for acceleration of charged particles which can be launched as black hole jets. 
Similarly, e.g. in the case of a uniform magnetic field considered above, the discharge of induced electric field $Q_{\bullet{\rm ind}}=2 a M_{\bullet} B$ would lead to the decrease of black hole spin $a$ and resulting extraction of rotational energy of a black
hole, while $B$ is constant by its definition.   

Multiple numerical simulations of two processes for general relativistic magnetohydrodynamic flow showing the efficient extraction of energy from rotating black holes imply that the induced electric field are not screened at least in the close vicinity of black holes. 
The problem of screening of induced electric field by surrounding plasma for Sgr~A* is discussed in Section~\ref{subsec_brems_hole}.

The value of $Q_{\bullet \rm ind}^{\rm max}$ is small in comparison with $Q_{\rm max}$ implying that its effect on the spacetime geometry can be neglected. The upper boundary for the charge-to-mass ratio for the Galactic centre SMBH \citep{2001PhRvD..63f4037K} is
\begin{align}
  \frac{Q_{\bullet \rm ind}}{M_{\bullet}}&=2 B_0 \frac{J_{\bullet}}{M_{\bullet}} \leq 2 B_0 M_{\bullet}=\,\notag \\
                                 &=8 \times 10^{-13} \left(\frac{B_0}{10\,{\rm G}}\right) \left(\frac{M_{\bullet}}{4 \times 10^6\,M_{\odot}}\right) \ll 1\,.
  \label{eq_wald_charge}                                
\end{align}
 Relation (\ref{eq_wald_charge}) implies that the induced black hole charge is weak in a sense that its effect on the dynamics of neutral matter can be neglected and Kerr metric approximation can be used. However, induced charge can have considerable effect on the dynamics of charged matter and consequently on some of the observables of Sgr~A*, which we specifically discuss in Section~\ref{section_observable_effects}.


%
%

\subsection{Charge fluctuation due to accretion}
\label{subsec_charge_accretion}

The electric charge of $Q_{\max}^{\rm norot}$ is a theoretical upper limit. In reality, the accretion of positively charged particles (protons) will stop to proceed when the Coulomb force between the SMBH and the proton is of the same value and opposite orientation as the gravitational force, giving the condition for the maximum positive charge in a non-relativistic case,
\begin{equation}
  Q_{\rm max}^{+}=4\pi \epsilon_0 G M_{\bullet}\left(\frac{m_{\rm p}}{e}\right)=6.16\times 10^8\,\left(\frac{M_{\bullet}}{4\times 10^6\,M_{\odot}} \right)\,C\,,
  \label{eq_max_plus_charge}
\end{equation}
which is much smaller than the maximum charge, $Q_{\rm max}^{+}/Q_{\rm max}=2\sqrt{\pi \epsilon_0 G}m_{\rm p}/e\approx 9\times 10^{-19}$.

In a similar way, the maximum negative charge derived for accreting electrons is,
\begin{equation}
  Q_{\rm max}^{-}=4\pi \epsilon_0 G M_{\bullet}\left(\frac{m_{\rm e}}{e}\right)=3.36\times 10^5\,\left(\frac{M_{\bullet}}{4\times 10^6\,M_{\odot}} \right)\,C\,,
  \label{eq_max_minus_charge}
\end{equation}
which leads to an even smaller ratio $Q_{\rm max}^{-}/Q_{\rm max}=2\sqrt{\pi \epsilon_0 G}m_{\rm e}/e\approx 4.9\times 10^{-22}$. The ratios of maximum positive and negative charges to the maximum charge allowed for the SMBH imply that the space-time metric is not affected by an electric charge in a significant way.

Eqs.~\eqref{eq_max_plus_charge} and \eqref{eq_max_minus_charge} are applicable only far from the black hole. In the relativistic regime, the motion of charged particles in the simplest regime without rotation can be studied within the Reissner-Nordström solution, which can be acquired from the Kerr-Newman metric by setting $a_{\bullet}=0$ in Eq.~\eqref{eq_kerr_newman}. The line element in the geometrized units with $c=1=G$ can be written as 
\beq 
d s^2 = - f(r) d t^2 + f(r)^{-1} d r^2 + r^2 d\theta^2 + r^2 \sin^2\theta d\phi^2 , \label{metric}
\eeq
where the lapse function is defined as
\beq 
f(r) = 1 - \frac{r_S}{r}  + \frac{r_Q^2}{r^2}. \label{lapsefunc}
\eeq
The four-vector potential $A_\alpha$ of electromagnetic field generated by the charge $Q_{\bullet}$ of the \RN~ black hole takes the form
\beq
A_{\mu} = \frac{Q_{\bullet}}{r} \delta^{(t)}_{\mu}.
\eeq
The dynamics of charged particle in curved spacetime in presence of electromagnetic fields is governed by the equation
\beq \label{eqmocp}
 \frac{d u^\mu}{d \tau} + \Gamma^\mu_{\alpha\beta} u^\alpha u^\beta = \frac{q}{m} F^{\mu}_{\,\,\, \nu} u^{\nu}, 
\eeq
where $u^{\mu} = d x^{\mu} / d\tau$ is the four-velocity of the particle with mass $m$ and charge $q$, normalized by the condition $u^{\mu} u_{\mu} = - 1$, ~$\tau$ is the proper time of the particle, $F_{\mu \nu} = A_{\nu,\mu} - A_{\mu,\nu}$ is an antisymmetric tensor of the electromagnetic field and components of $\Gamma^\mu_{\alpha\beta}$ are the Christoffel symbols.

One of the interesting features of the motion of charged particles in the vicinity of \RN~ black hole is the existence of trapped equilibrium state of a particle where the electrostatic forces between two charges compensate the gravitational attraction of a black hole. The four velocity of a particle at this state is 
$u^{\alpha} = \left(1/\sqrt{- g_{tt}}, 0, 0, 0\right)$. Thus, the time component of equation (\ref{eqmocp}) takes the form
\beq \label{eq-for-Q-equil}
r_S - 2 \frac{r_Q^2}{r} + 2 \frac{Q q}{m} \left(1-\frac{r_S}{r} - \frac{r_Q^2}{r^2}\right)^{1/2} = 0 
\eeq
Since the gravitational effect of the charge $Q_{\bullet}$ is small in comparison with those of the mass of the black hole (corresponding to $r_Q \ll r_S$) one can neglect the higher order terms in $Q_{\bullet}$ equation (\ref{eq-for-Q-equil}). Thus, for the charge $Q$ we get
\beq \label{Q-equil}
Q_{\bullet} = \frac{m r_S}{2 q} \left(1-\frac{r_S}{r}\right)^{-1/2}.
\eeq
The charge (\ref{Q-equil}) can be interpreted as the maximum net charge that can be accreted into the black hole from the given position $r$ before the electrostatic force will prevail and the accretion of same-charge particles stops. Restoring the constant in (\ref{Q-equil}) we get charge in SI units as
\beq \label{Q-equil-SI}
Q_{\rm max}^{\rm rel} = 4 \pi \epsilon_0 G M_{\bullet} \frac{m_{\rm par}}{q_{\rm par}} \left(1-\frac{r_S}{r}\right)^{-1/2}.
\eeq
The factor $\left(1-\frac{r_S}{r}\right)^{-1/2}$ is the general relativistic correction to the corresponding Newtonian equation. This implies that the electrostatic force increases while approaching black hole. Close to the horizon, the divergence of (\ref{Q-equil-SI}) means that the black hole requires infinite charge in order to keep the equilibrium position of the particle. We plot the ratio $Q_{\rm max}^{\rm rel}/Q_{\rm max}^{+/-}$, i.e. the relativistic correction in Fig.~\ref{fig_rel_correction}.

\subsection{Charge and dynamical timescales}
The electric charge of the SMBH is expected to fluctuate due to discharging by particles of an opposite charge. This is especially efficient when the free-fall timescale of particles with the opposite charge is significantly shorter than the free-fall timescale of the particles with the same charge as that of the black hole. The free-fall timescale for a charged black hole is modified due to the presence of an additional Coulomb term in the equation of the motion for a radial infall (neglecting the gas pressure),

\begin{equation}
   \frac{\mathrm{d}v}{\mathrm{d}t}=-\frac{GM_{\bullet}}{r^2}+\frac{1}{4\pi\epsilon_0}\frac{Q_{\bullet}}{r^2}\frac{q_{\rm par}}{m_{\rm par}}\,,
   \label{eq_motion_charge}
\end{equation} 
where $q_{\rm par}$ and $m_{\rm par}$ are the charge and the mass of the particle, respectively. The signs in the equation are the following: for the positive charge of the black hole $Q_{\bullet}=+Q^{+}$, the particle charge is $q_{\rm par}=+e$ for the proton $(m_{\rm par}=m_{\rm p})$ and $q_{\rm par}=-e$ for the electron $(m_{\rm par}=m_{\rm e})$. For the negative charge of the black hole $Q_{\bullet}=-Q^{-}$, the signs of particle charges are kept as before.

The Newtonian free-fall timescale derived from Eq.~\ref{eq_motion_charge} for a particle falling in from the initial distance of $r_0$ is

\begin{equation}
  t_{\rm ff,Q}=\frac{\pi r_0^{3/2}}{\sqrt{8(GM_{\bullet}-\frac{1}{4\pi\epsilon_0} Q_{\bullet}\frac{q_{\rm par}}{m_{\rm par}}} ) }\,,
  \label{eq_free_fall_charge}
\end{equation}
which becomes $t_{\rm ff}(Q_{\bullet}=0)=\pi r_0^{3/2}/\sqrt{8GM_{\bullet}}$ for zero charge of the SMBH. The direct outcome of Eq.~\eqref{eq_free_fall_charge} is the difference of free-fall timescales for protons and electrons for a given charge of the SMBH. 

\begin{figure*}
 \centering
 \begin{tabular}{cc}
   \includegraphics[width=0.5\textwidth]{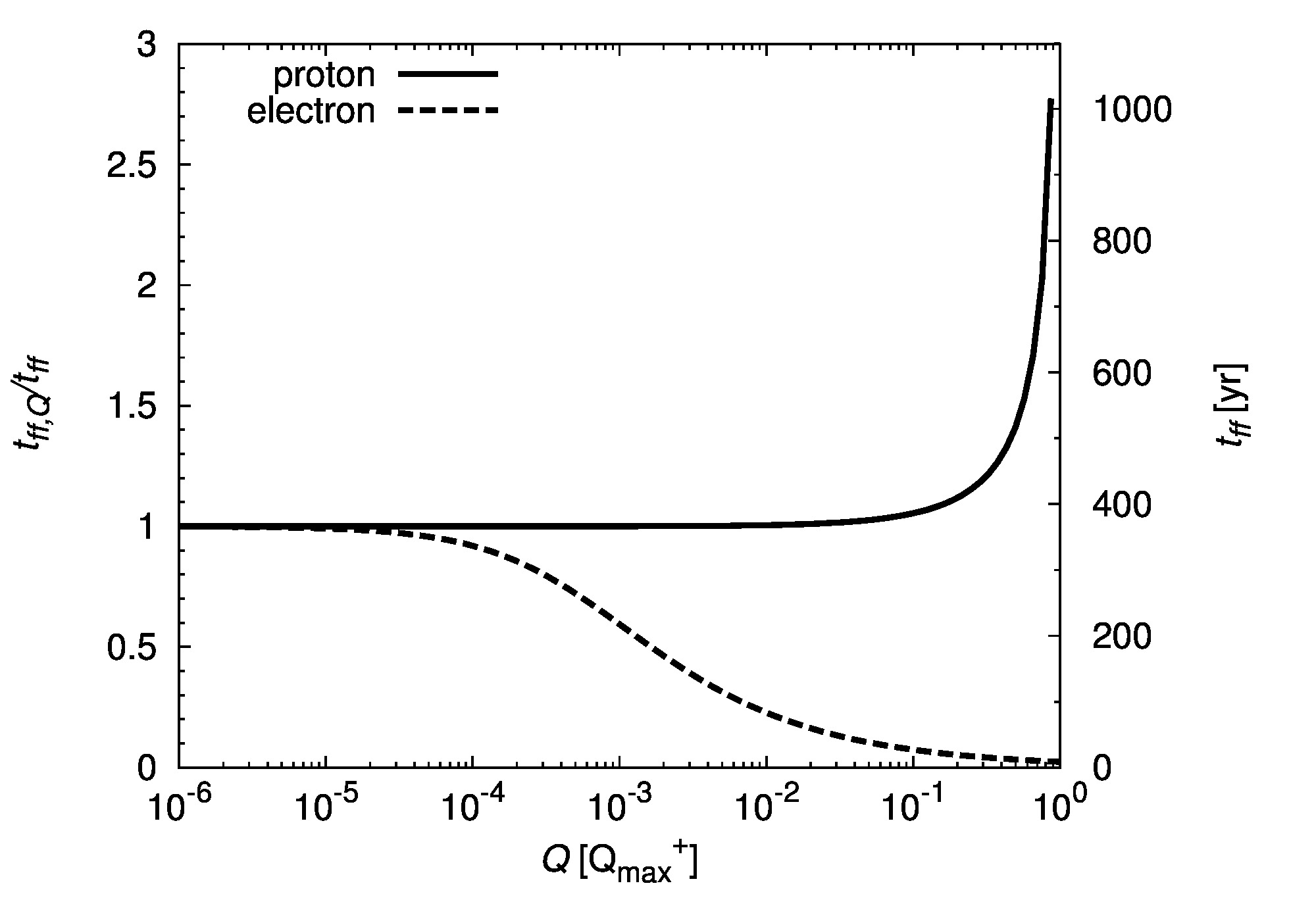} & \includegraphics[width=0.5\textwidth]{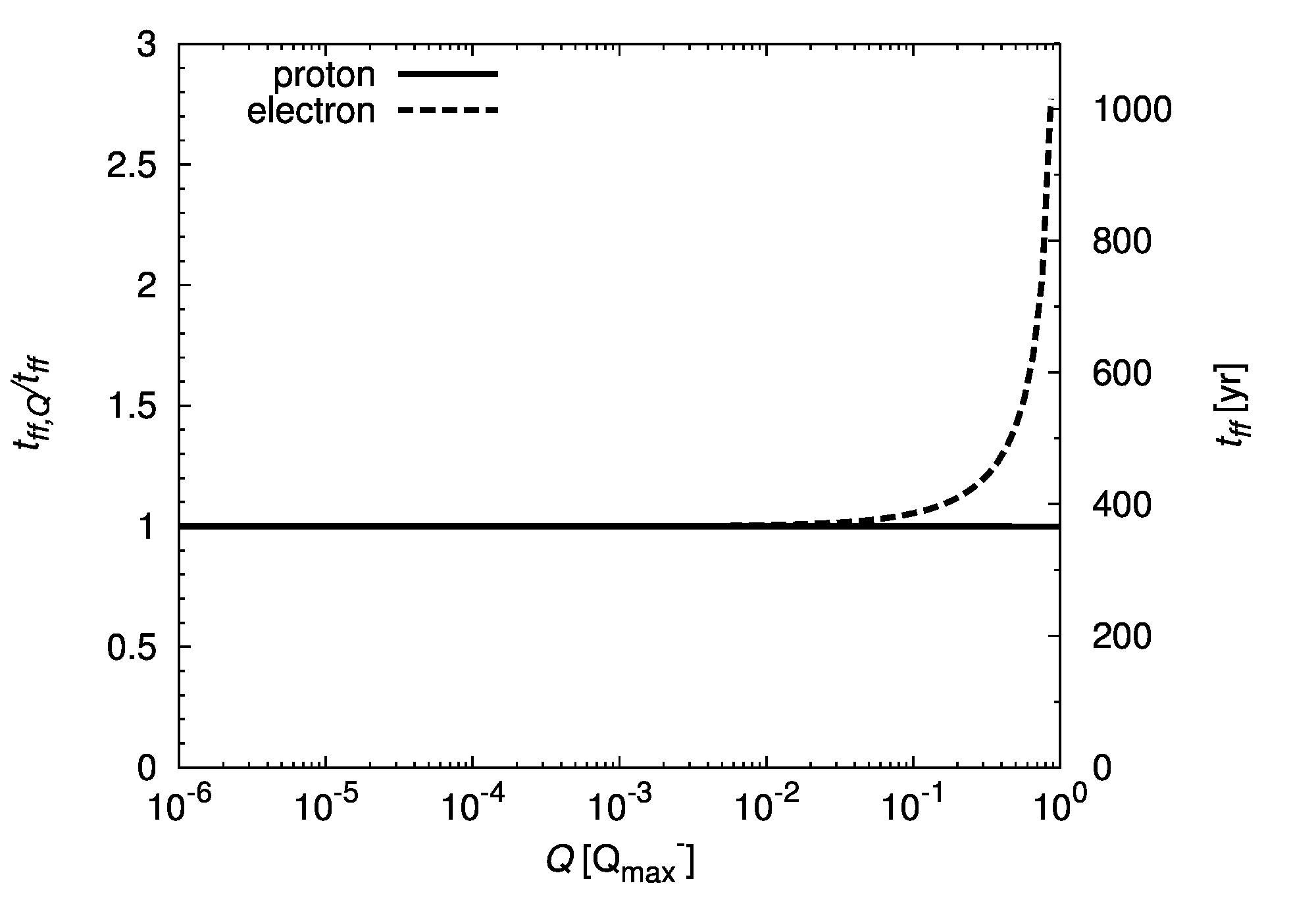}\\
 \end{tabular}
 \caption{\textbf{Left panel:} The free-fall timescales for protons and electrons calculated for a positive charge of the supermassive black hole at the Galactic centre expressed in units of the maximum positive charge ($Q_{\rm max}^{+}$, see Eq.~\ref{eq_max_plus_charge}). The scale on the left vertical axis expresses the ratio $t_{\rm ff,Q}/t_{\rm ff}$ whereas the scale along the right vertical axis is the free-fall timescale for an initial distance equal to Bondi radius (Eq.~\eqref{eq_Bondi_radius}) expressed in years. \textbf{Right panel:} The same as in the left panel, calculated for a negative charge of the SMBH expressed in units of the maximum negative charge ($Q_{\rm max}^{-}$, see Eq.~\ref{eq_max_minus_charge}).}
 \label{fig_freefall_timescales} 
\end{figure*}

In Fig.~\ref{fig_freefall_timescales}, we plot the free-fall timescales for protons and electrons for a positive charge of the SMBH (left panel) and for its negative charge (right panel). The timescales are comparable up to $Q^{+}\lesssim 10^{-5}\,Q_{\rm max}^{+}$ for the positively charged black hole and up to $Q^{-} \lesssim 10^{-2}\,Q_{\rm max}^{-}$, which further limits a charge of the SMBH since for larger charges the infall of opposite charges is progressively faster than the infall of the same charges. For the maximum positive charge of Sgr~A*, the free-fall timescale for electrons is $t_{\rm ff,Q_{\rm max}^{+}}=8.5\,{\rm yr}$ for an initial distance at the Bondi radius. This is a much shorter timescale than the free-fall timescale of protons for the maximum negative charge, which is close to the free-fall timescale for a non-charged black hole with an initial distance at the Bondi radius, $t_{\rm ff}=366\,{\rm yr}$. 

The free-fall time-scale from the Bondi radius is the basic dynamical timescale in the accretion flow. Any disturbance in the accretion flow develops on the viscous timescale given by Eq.~\eqref{eq_viscous_timescale}, which for the assumption of the thick flow $H\approx r_0$ and $\alpha\approx 0.1$ is approximately $t_{\rm vis}\approx 10 t_{\rm ff}(r_0,Q_{\bullet})$. 

In the following, we define a specific charge of accreted matter $\epsilon$, which relates the accretion rate of the charged matter to the total accretion rate as $\dot{M}_{\rm acc}^{\rm charge}=\epsilon \dot{M}_{\rm acc}$. From this relation, the charging rate of the black hole, $\dot{Q}_{\bullet}$, may be expressed as,

\begin{equation}
  \dot{Q}_{\bullet}=\epsilon \frac{q_{\rm par}}{m_{\rm par}}\dot{M}_{\rm acc}\,,
  \label{eq_charge_rate}
\end{equation}
where $\dot{M}_{\rm acc}$ is the total accretion rate. The total accretion rate was inferred from the observations via the Faraday rotation by \citet{2007ApJ...654L..57M}, who obtain $\dot{M}_{\rm acc}=2\times 10^{-9}\,{\rm M_{\odot}\,yr^{-1}}$ up to $2\times 10^{-7} {\rm M_{\odot}\,yr^{-1}}$, depending on the configuration of the magnetic field. For the induced positive charge  $Q_{\rm ind}^{+}$, the charging (induction) time-scale follows from Eq.~\eqref{eq_charge_rate},

\begin{equation}
  \tau_{\rm charge}^{+}=\frac{m_{\rm p} Q_{\rm ind}^{+}}{e\epsilon_{\rm pos}\dot{M}_{\rm acc}}\,,
 \label{eq_discharge_max_plus} 
\end{equation}
while for the induced negative charge we get,
\begin{equation}
  \tau_{\rm charge}^{-}=\frac{m_{\rm e} Q_{\rm ind}^{-}}{e\epsilon_{\rm neg}\dot{M}_{\rm acc}}\,.
 \label{eq_discharge_max_minus} 
\end{equation}

To charge the black hole positively, mainly by the induction process described in Section~\ref{subsec_rotating_SMBH}, the charging timescale expressed by Eq.~\eqref{eq_discharge_max_plus} needs to be smaller than the discharge timescale, which can be estimated by the viscous timescale of electrons on the scale of the gravitational radius. On the other hand, the charging timescale must be larger than the timescale given by the plasma frequency for protons, $\tau_{\rm p}=2\pi(\epsilon_0 m_{\rm e}/(n_{\rm e}e^2))^{1/2}$, which expresses the charged density fluctuations on scales larger than the Debye length. For the exemplary values of $Q_{\rm ind}=Q_{\bullet,10}10^{10}\,C$ and $\dot{M}_{\rm acc}=\dot{M}_{-8}=10^{-8}\,M_{\odot}{\rm yr^{-1}}$, we obtain the limits

\begin{equation}
1.3\times 10^{-13} Q_{\bullet,10}\dot{M}_{-8}^{-1} \lesssim \epsilon_{\rm pos} \lesssim 1.4 \times 10^{-6}Q_{\bullet,10}\dot{M}_{-8}^{-1} \,.
 \label{eq_constraints_positive}
\end{equation}

In an analogous way, the negative charged fraction of the accretion rate is
\begin{equation}
1.7\times 10^{-18}Q_{\bullet,10}\dot{M}_{-8}^{-1} \lesssim \epsilon_{\rm neg} \lesssim 3.3 \times 10^{-8}Q_{\bullet,10}\dot{M}_{-8}^{-1}\,,
 \label{eq_constraints_negative}
\end{equation}
as inferred by comparing \eqref{eq_discharge_max_minus} to the viscous timescale of protons (upper timescale limit for discharging) and to the timescale of electron plasma oscillations (both evaluated at the ISCO scale of $r_{\rm ISCO}\approx GM_{\bullet}/c^2$).

The charging of Sgr~A* can thus effectively proceed when a rather small fraction of accreted matter ($\epsilon_{\rm pos}$ or $\epsilon_{\rm neg}$) is charged. The charging process of the accreted fluid may proceed at a larger distance from the SMBH horizon plausibly due to strong irradiation or the external (Galactic) magnetic field \citep{2011PhRvD..84h4002K}.

\begin{figure*}
  \centering
  \includegraphics[width=0.45\textwidth]{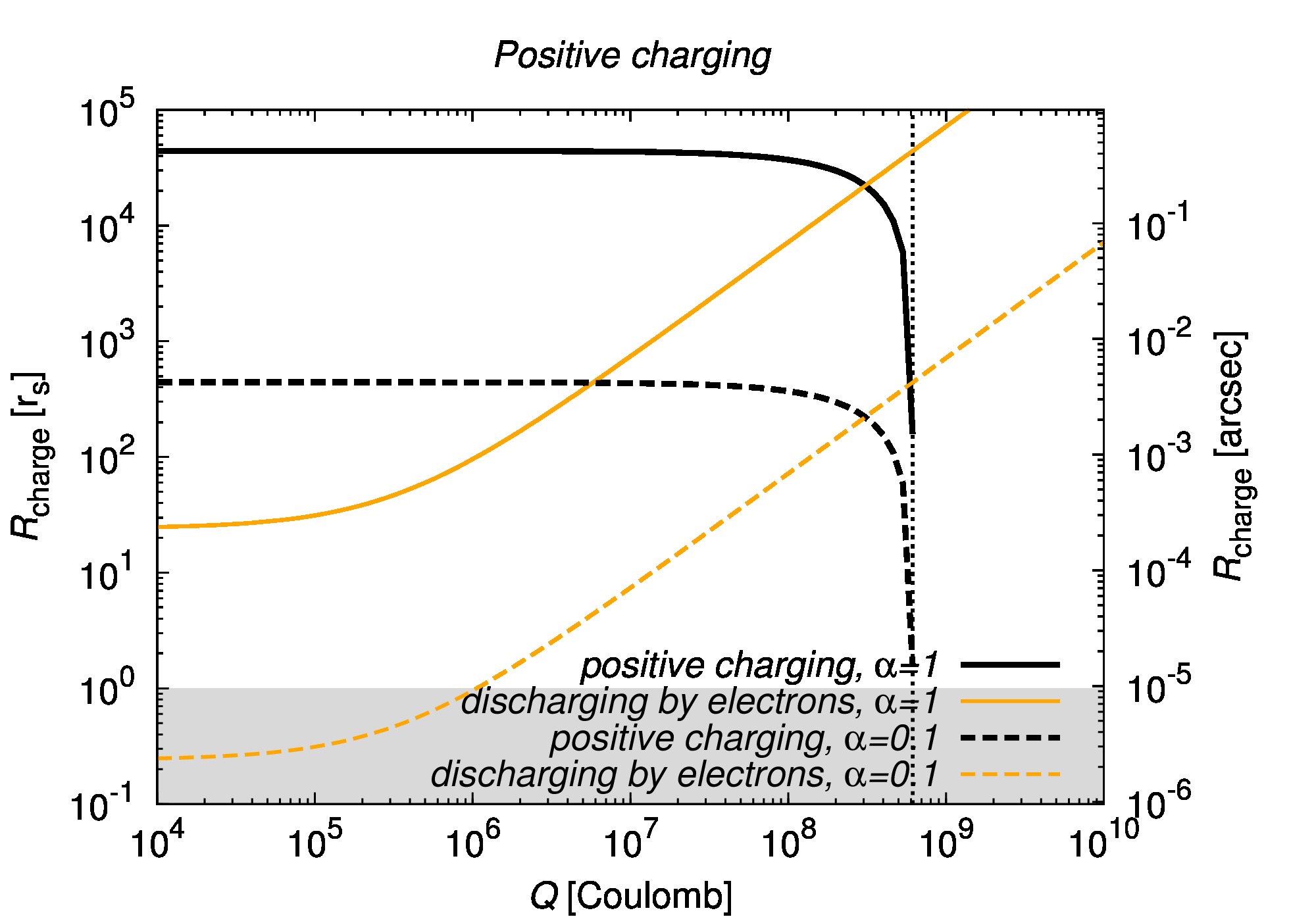}
    \includegraphics[width=0.45\textwidth]{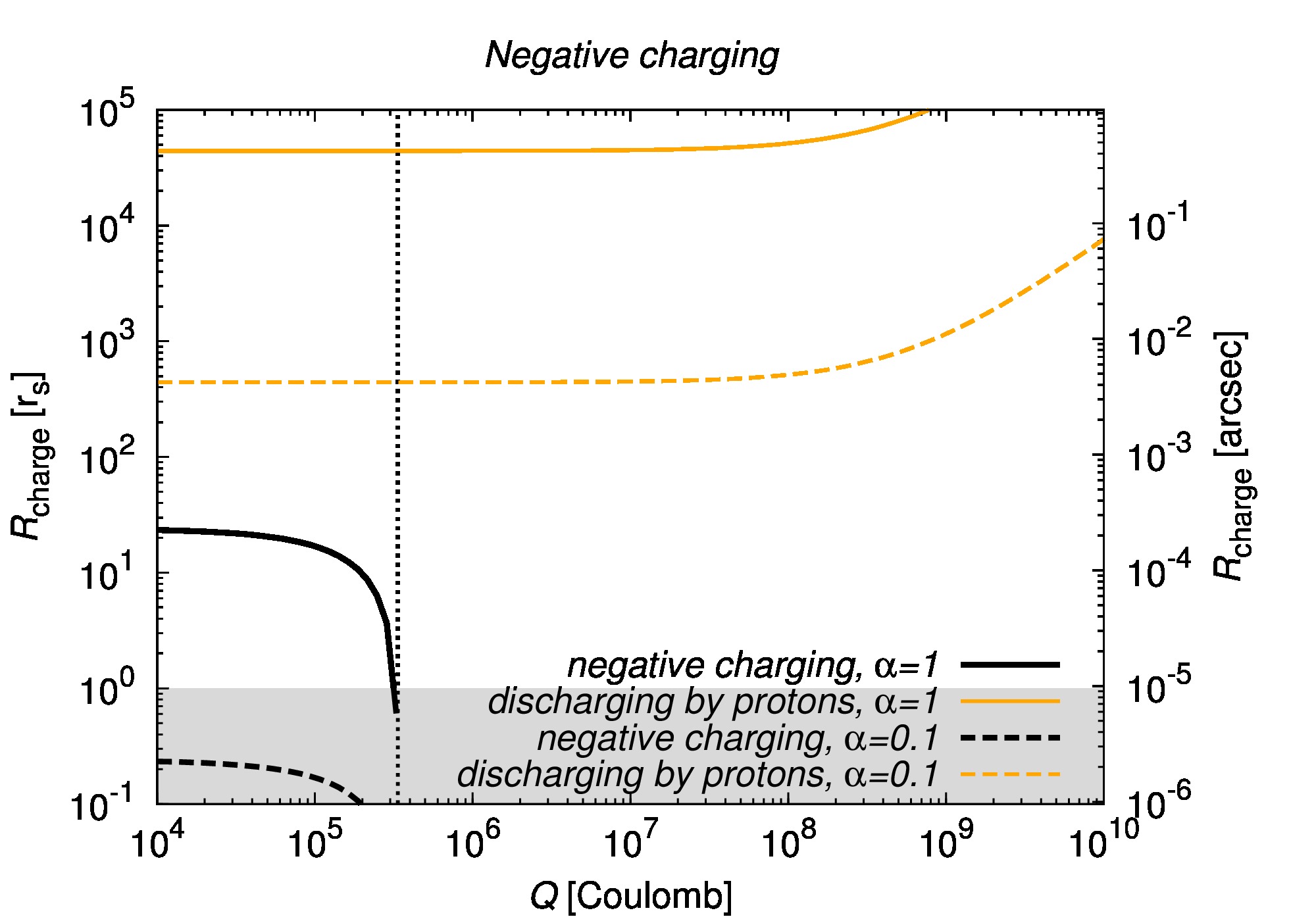}
  \caption{The dependence of the radius of the sphere, where the basic condition for the charging or the discharging, $t_{\rm vis, Q}<t_{\rm th}$, is met for both the positive charging (left panel) and the negative charging (right panel). The solid lines mark the charging/discharging length-scale for the purely free-fall accretion while the dashed lines represent the viscous infall for $\alpha=0.1$. The charging can effectively proceed when the charging length-scale $R_{\rm charge}$ is larger than the discharging length-scale. The shaded region marks the region below the event horizon.}
  \label{fig_radius_charging}
\end{figure*} 

The process of the black hole charging can proceed efficiently when the gravitational and electrostatic forces acting on a particle are greater than the thermal pressure forces inside a region of radius $R_{\rm charge}$, which is equivalent to the condition that the viscous time-scale of the inward motion must be smaller than the thermal time-scale, $t_{\rm vis, Q}<t_{\rm th}$ inside $R_{\rm charge}$. The thermal time-scale is simply, $t_{\rm th}=R_{\rm charge}/v_{\rm th}$, where the thermal speed $v_{\rm th}$ is related to either electrons or protons. The condition of the smaller viscous time-scale is met inside the sphere of radius $R_{\rm charge}$, whose radius progressively gets smaller and it reaches zero at either the maximum positive or negative charge, see Eqs.~\eqref{eq_max_plus_charge} and \eqref{eq_max_minus_charge}. Under the assumption that the accretion flow is thick in a sense $(H/r_0)\approx 1$, it can be simply derived that,

\begin{equation}
  R_{\rm charge}\lesssim \left(\alpha/\pi\right)^2 \frac{m_{\rm par}}{k_{\rm B} T_{\rm g}} (8GM_{\bullet}-\frac{2}{\pi \epsilon_0}Q_{\bullet} \frac{q_{\rm par}}{m_{\rm par}})\,,
  \label{eq_r_charge}
\end{equation}
with the sign convention as in Eq.~\eqref{eq_free_fall_charge}. For the zero charge of the black hole, the relation~\eqref{eq_r_charge} is similar to the definition of the Bondi radius, Eq.~\eqref{eq_Bondi_radius}, $R_{\rm charge} \lesssim 8(\alpha/\pi)^2 GM_{\bullet}/v_{\rm th}^2$.

In Fig,~\ref{fig_radius_charging}, we plot Eq.~\eqref{eq_r_charge} for both positive charging (protons falling into the positively charged black hole; see left panel) and negative charging (electrons falling into negatively charged black hole; right panel). Inside the radius $R_{\rm charge}$ basic condition for charging is met -- particles with the same sign of the charge are not prevented from descending towards the black hole by thermal pressure. However, the particles of the opposite charge also fall in progressively faster inside the discharging sphere with the radius of $R_{\rm discharge}$ defined analogously to Eq.~\eqref{eq_r_charge}, which effectively limits the realistic values of the electrostatic charge of the black hole.

The necessary condition for an increasing charge of the SMBH is that $R_{\rm charge} \gtrsim R_{\rm discharge}$, which is only met for the positive charging, see the left panel of Fig.~\ref{fig_radius_charging}. For negative charging, the discharging length-scale of protons is always larger than the charging length-scale of electrons purely because of the mass difference. Therefore, the negative charging of black holes is rather inefficient. 

Hence, the Galactic centre black hole and black holes in general are prone to have a small positive charge. This is also supported by the analysis in Section~\ref{subsec_rotating_SMBH}, where it is shown that black holes whose spin is oriented parallel to the magnetic field intensity preferentially accrete positively charged particles, while those with anti-parallel spin are being negatively charged. Since we expect a certain degree of alignment in a relaxed system of a black hole and its associated accretion flow, the induced charge is expected to be positive.

For the positive charging in Fig.~\ref{fig_radius_charging} (left panel), the charging length-scale for protons is larger than the discharging radius of electrons up to a certain charge $Q_{\rm eq}$, which for the general case of different electron and proton temperatures is,

\begin{equation}
  Q_{\rm eq}=\frac{4\pi \epsilon_0 GM_{\bullet}}{e}\frac{[(T_{\rm e}/T_{\rm p})m_{\rm p}-m_{\rm e}]}{1+T_{\rm e}/T_{\rm p}}\,.
  \label{eq_charge_rcharge}
\end{equation}
For the case of the same temperature $T_{\rm e}\approx T_{\rm p}$, Eq.~\eqref{eq_charge_rcharge} becomes identical to the equilibrium charge in Eq.~\eqref{eq_charge_blackhole}, $Q_{\rm eq}=3.1 \times 10^8 \left(M_{\bullet}/4\times 10^6\,M_{\odot} \right)\,C$. This charge is associated with the charging/discharging length-scale of $R_{\rm charge}=0.2''$ for the free-fall accretion and $R_{\rm charge}=2.1\,{\rm mas}$ for the accretion with the viscosity parameter of $\alpha=0.1$, which is one and three orders of magnitude smaller than the Bondi radius, respectively.

\section{Observable effects associated with a charged black hole}
\label{section_observable_effects}

Although the simple analysis based on the first principles showed that the charge associated with the Galactic centre black hole is at least ten orders of magnitude smaller than the charge corresponding to an extremal black hole, it is useful to list observational 
signatures that a charged black hole could have. The electric charge, if present, most likely does not reach values significant for the spacetime metric. However, despite its expected small value, it is useful to design observational tests of its presence. It is therefore of astrophysical interest to name several potential observables that can be employed to test the presence of the charged SMBH at the Galactic centre.

\subsection{Effect on the black hole shadow}
\label{effect_shadow}

 For shorter wavelengths in the radio domain than $\lambda_{\rm p}$, plasma does not block the emission, however it causes significant scatter broadening of any structure up $1.4\,{\rm mm}$ \citep{1998ApJ...508L..61L}. At wavelengths $\lambda<1.4\,{\rm mm}$, it is possible to resolve the structure using the VLBI, since the size of Sgr~A* starts to be source-dominated \citep{1998A&A...335L.106K,2008Natur.455...78D}. Of a particular interest is a well-defined curve on the sky plane which divides the region where photon geodesics intersect the even horizon from the region where photons can escape to the observer --  the so-called black hole shadow \citep{2000ApJ...528L..13F}.  

It was previously claimed by \citet{2014PhRvD..90f2007Z} that the charge associated with the Galactic centre black hole could be detected via the VLBI imaging of Sgr~A*, based on the detection of the shadow. They report that a Reissner-Nordström black with a significant charge close to $Q_{\bullet}=Q_{\rm max}^{\rm norot}$ is more consistent with the VLBI detection of Sgr~A* by \citet{2008Natur.455...78D} with the core diameter of $37^{+16}_{-10}\,\mu {\rm as}$. This argument is based on the theoretical calculations of the shadow diameter, namely the shadow diameter for the Schwarzschild black hole is $6\sqrt{3} GM_{\bullet}/c^2$, which is $\sim  
51.2\,\mu{\rm as}$ at the distance of $8\,{\rm kpc}$ at the Galactic centre. For an extremal Reissner-Nordström black hole, one gets the shadow diameter $4 GM_{\bullet}/c^2$, i.e. by about $38\%$ smaller than for the Schwarzschild black hole, which corresponds to $\sim 39.4\,\mu{\rm as}$. The shadow diameter for the nearly extremal charged black hole is thus closer to the core size found by \citeauthor{2008Natur.455...78D}~(\citeyear{2008Natur.455...78D}).

\begin{figure*}
 \centering
 \begin{tabular}{cc}
   \includegraphics[width=0.5\textwidth]{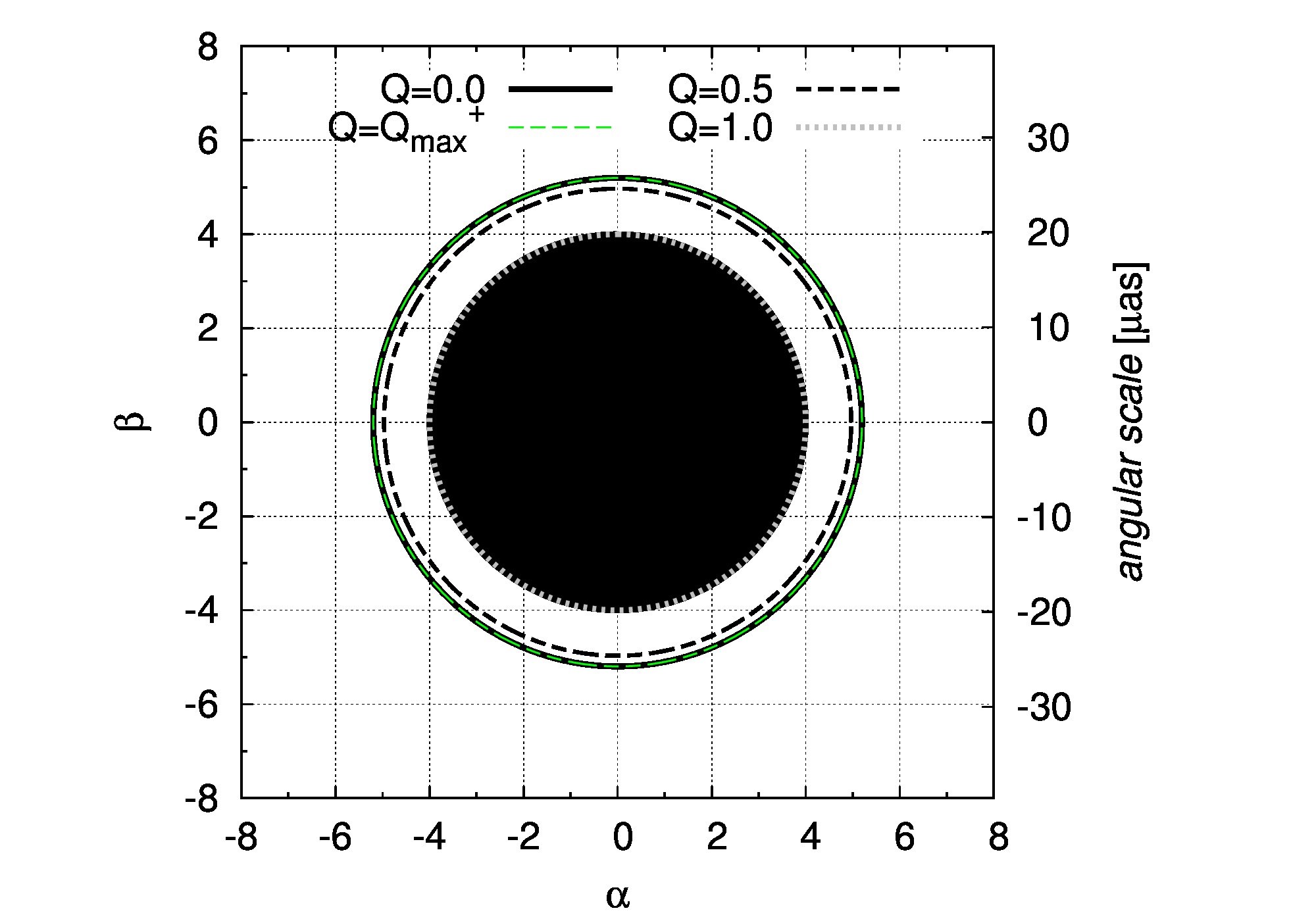}         \includegraphics[width=0.5\textwidth]{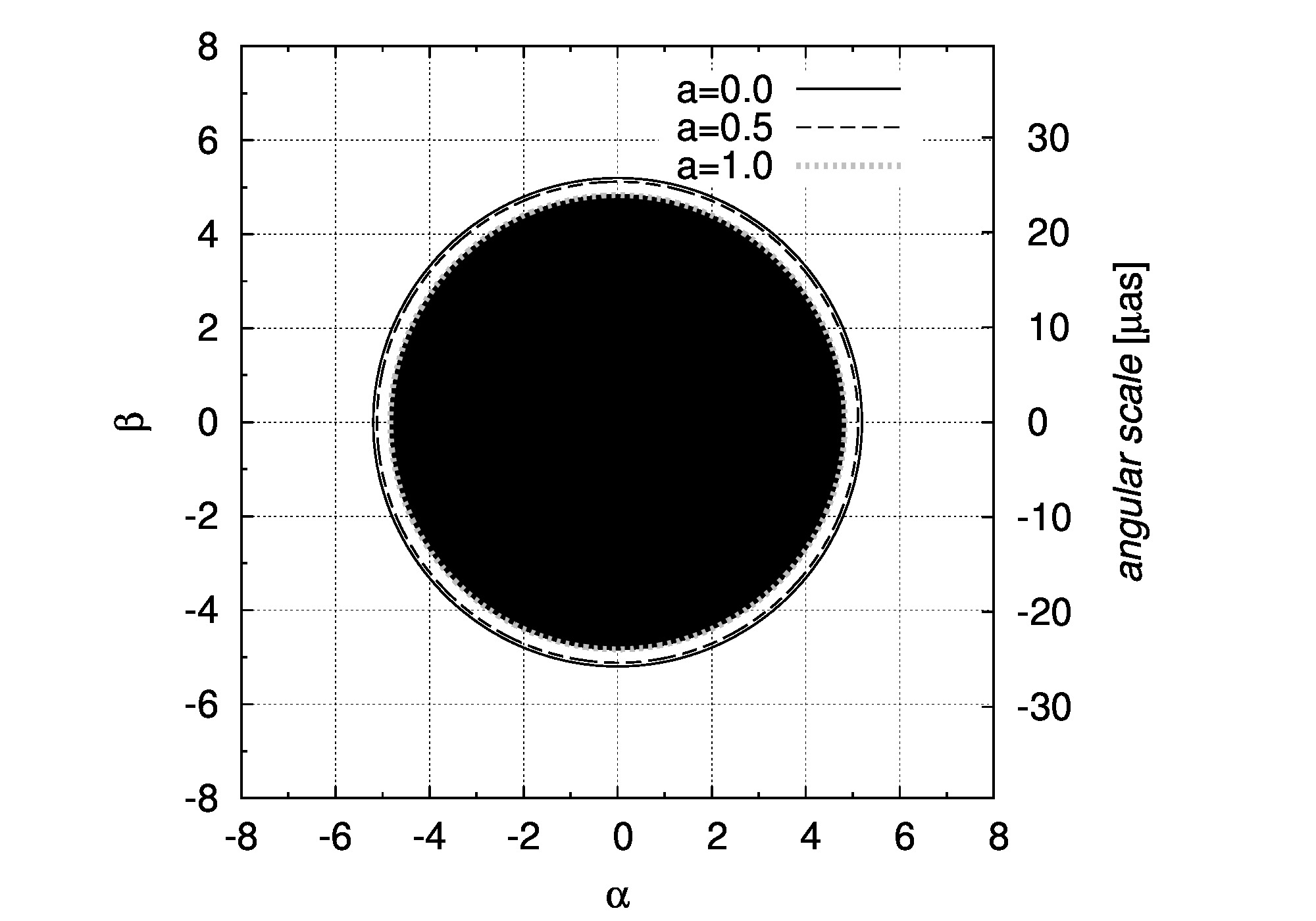}
 \end{tabular}
 \caption{\textbf{Left:} The black hole shadow size in gravitational units for a non-rotating, Reissner-Nordström black hole for an increasing value of the charge according to the legend. The right vertical axis is expressed in microarcseconds for the distance of the Galactic centre -- $8\,{\rm kpc}$. \textbf{Right:} The same as the left figure but for the rotating black hole (Kerr). The calculations are performed for the top view, in the direction of the spin axis, for increasing values of the spin parameter according to the legend.}
 \label{fig_shadow_size}
\end{figure*}

However, the core size found by \citet{2008Natur.455...78D} does not correspond in a straightforward way to the black hole shadow size. It can be either a Doppler beamed accretion flow or the footpoint of the jet \citep{2008Natur.455...78D,2012MNRAS.421.1517D,2017FoPh...47..553E} and in these cases it is difficult to make a connection to the charge of the black hole. In general, the black hole shadow is not a clean observable. Not only does the charge influence its size, but also the spin, see Fig.~\ref{fig_shadow_size} for comparison. In addition, the charge starts significantly influencing the size of the shadow for fractions of the maximum charge, $Q_{\bullet} \gtrsim 0.1\,Q_{\rm max}^{\rm norot}$. For the maximum positive charge value of $Q_{\rm max}^{+}\approx 6 \times 10^8\,C$, see Eq.~\eqref{eq_max_plus_charge}, there is practically no difference in the size of the shadow.

Below we show that the bremsstrahlung surface brightness profile on the scales of $100\,r_{\rm S}$ up to $1000\,r_{\rm S}$ can be used to test the presence of a much smaller charge associated with Sgr~A* than by using the shadow size.

\subsection{Testing the presence of a Black Hole Debye shield -- does a charged black hole have an impact on the bremsstrahlung profile?}
\label{subsec_brems_hole}

The charge associated with Sgr~A* can have a considerable impact on the motion and the distribution of charged particles, electrons, protons, and ions, in its vicinity. However, this only applies to the charged black hole that is not shielded. In the classical plasma theory, any charged body immersed in stationary plasma with the electron density $n_{\rm e}$ and temperature $T_{\rm e}$ is shielded beyond the characteristic Debye length-scale, $\lambda_{\rm D}=\sqrt{\epsilon_0 k_{\rm B} T_{\rm e}/(n_{\rm e} e^2)}$, which results in the exponential potential decrease, $\phi=\phi_0\exp{(-r/\lambda_{\rm D})}$, where $\phi_0$ is the potential of a point charge in vacuum. 

The plasma around Sgr~A* is, however, certainly not stationary but rather dynamic, given the large velocity of accretion flow in the potential of Sgr~A* and turbulence. Therefore, the standard Debye theory is not applicable to this environment.

Even if the Debye shield around Sgr~A* were created, it would have such a small length-scale that it would completely lie inside the ISCO, where it would be dynamically sheared and it would be therefore highly unstable. When evaluated on the scale of the ISCO, using the extrapolated density and temperature profiles in Eqs.~\ref{eqs_powerlaws_bondi}, the Debye length is,
\begin{equation}
  \lambda_{\rm ISCO}=5\,\left(\frac{T_{\rm e}}{8.7\times 10^{12}\,{\rm K}} \right)^{1/2}\left(\frac{n_{\rm e}}{1.7\times 10^9\,{\rm cm^3}} \right)^{-1/2}\,{\rm m}\,,
  \label{eq_Debye_Bondi}
\end{equation}
while at the Bondi radius it would be only one order of magnitude larger, $\lambda_{\rm Bondi}\approx 141\,{\rm m}$.

 Moreover, using the classical estimates, the Debye sphere would not be formed based purely on viscous timescales if the charged fraction of accreted matter is large enough. If we imagine that the positively charged black hole is surrounded by a negatively charged Debye shell, its mass can be estimated simple as $M_{\rm Debye}\approx (Q_{\rm ind}^{+}/e)m_{\rm e}$, where $Q^{+}_{\rm ind}$ is the induced positive black hole charge. Since the Debye shell lies inside the ISCO, it is being depleted as well as refilled all the time. If the depletion rate of electrons $\dot{M}^{-}_{\rm acc}=\epsilon_{\rm neg}\dot{M}_{\rm acc}$ is larger than the filling rate $\dot{M}_{\rm Debye}$, then the Debye shell is highly transient and does not screen out the charge of the black hole. The filling rate of the Debye shell is assumed to take place on the viscous timescale of electrons, taking into account the positive charge of the black hole, $t_{\rm vis}^{-}=\alpha^{-1}(H/r_{\rm ISCO})^{-2}t_{\rm ff}(r_{\rm ISCO},Q_{\rm ind}^{+})$. Then the Debye filling rate can be expressed as,

\begin{equation}
 \dot{M}_{\rm Debye}\sim \frac{Q_{\rm ind}^{+}m_{\rm e}}{e t_{\rm vis}^{-}}\,.
 \label{eq_debye_filling}
\end{equation}
The Debye shell will not form if $\dot{M}^{-}_{\rm acc}>\dot{M}_{\rm Debye}$, which puts the lower limit on the negatively charged fraction of the accreted matter, $\epsilon_{\rm neg} > Q_{\rm ind}^{+}m_{\rm e}/(e \dot{M}_{\rm acc} t_{\rm vis}^{-}) \approx 7 \times 10^{-17}$. It implies that if the accreted matter contains a negatively charged fraction of the order of $\epsilon_{\rm neg}$, then the Debye shell is expected not to form at all.  Even if the Debye shell forms temporarily, it would be strongly perturbed by the turbulence and outflows. In addition, a more general analysis done by \citet{1978ApJ...220..743B} showed that all self-gravitating systems with length-scales $L$ larger than the Debye-length, $L\gg \lambda_{\rm D}$, are positively charged in order to hold in the electron gas, i.e. these objects are not Debye-screened.

An unshielded charged black hole would lead to the charge separation that directly influences the emissivity of the thermal bremsstrahlung, since the emission efficiency drops significantly for like particles, proton-proton and electron-electron interactions, since there is no dipole component in the collisions. The bremsstrahlung is dominantly produced by radiating electrons that move in the Coulomb field of protons (or positively charged ions) and the corresponding emissivity is given by Eq.~\eqref{eq_brems_lum}, $\epsilon_{\rm Brems} \propto Z^2 n_{\rm i} n_{\rm e}$.  Therefore, inside the sphere of the electrostatic influence of the black hole, the drop in bremsstrahlung emissivity is expected, creating a drop or ``hole'' in the surface density of the thermal bremsstrahlung, which is centred at Sgr~A*.

\begin{figure*}
  \centering
  \begin{tabular}{cc}
  \includegraphics[width=0.5\textwidth]{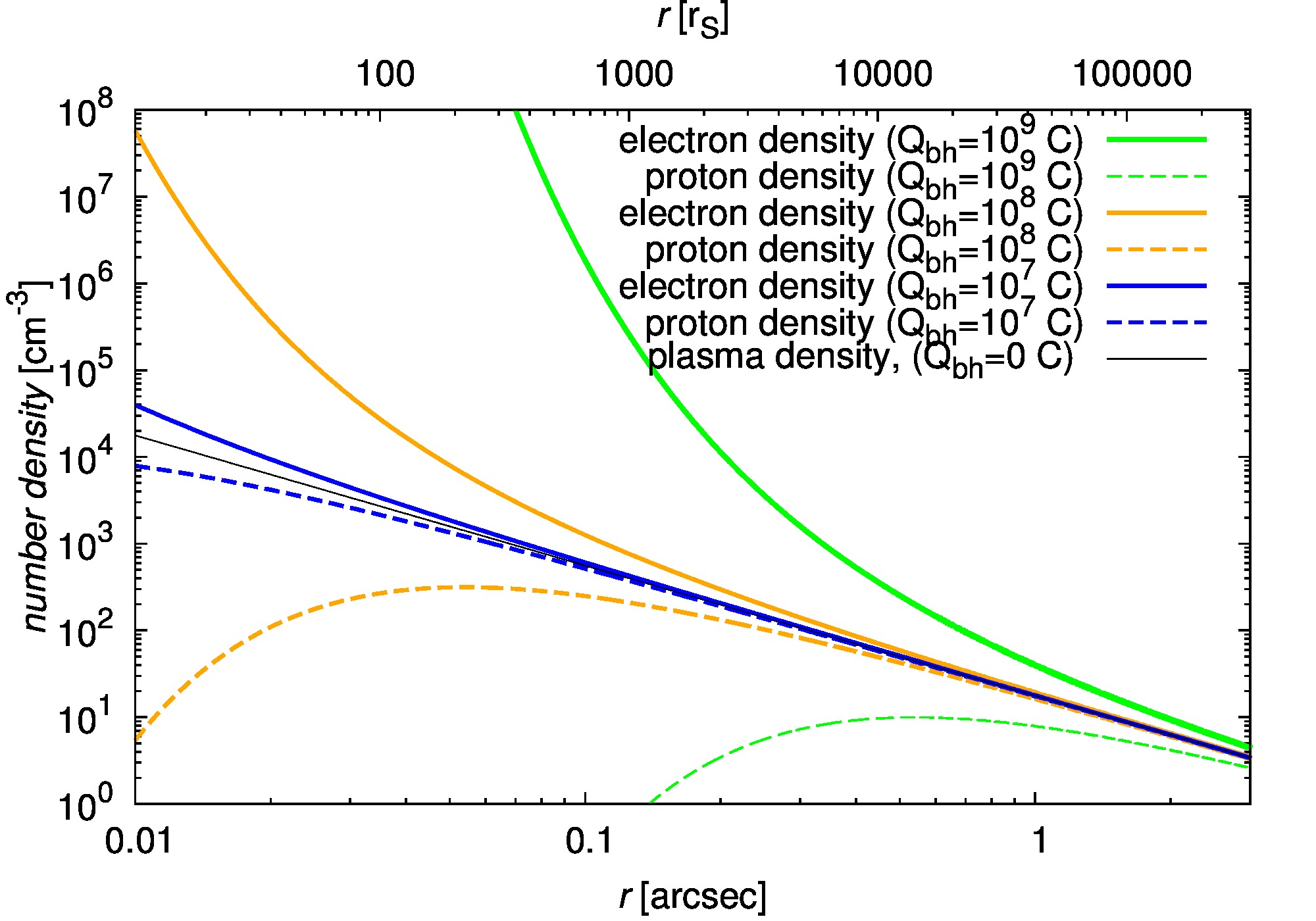} & \includegraphics[width=0.5\textwidth]{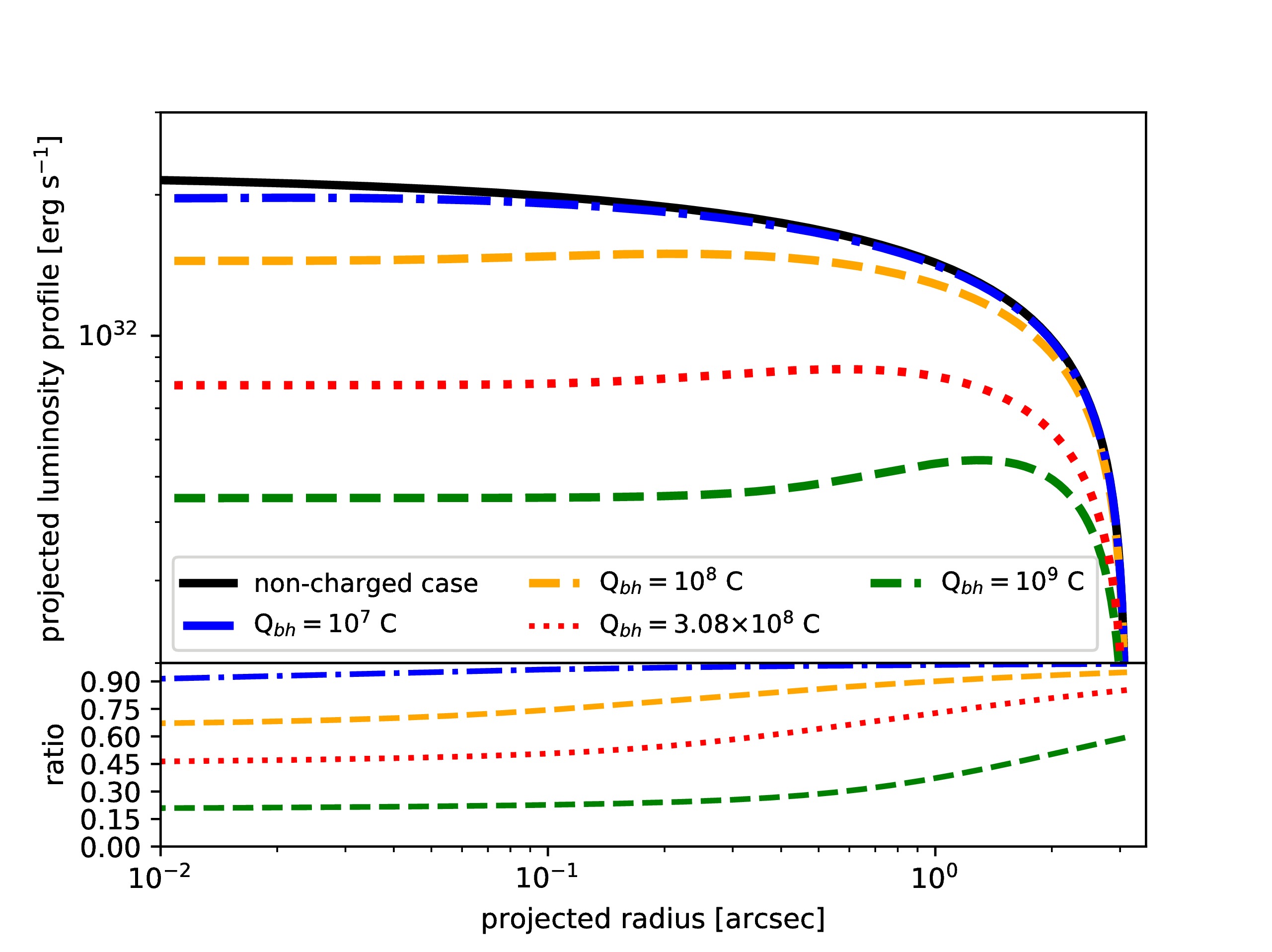}
  \end{tabular}
  \caption{\textbf{Left:} Density profiles of electrons and protons in the Galactic centre as affected by a potential charged supermassive black hole. The black solid line represents the non-charged power-law. The coloured lines (blue, orange, green) manifest the changed profiles due to the presence of a positive point charge at the centre: the blue solid and dashed lines represent the case for $Q_{\bullet}=10^7\,{\rm C}$, the orange lines stand for $Q_{\bullet}=10^8\,{\rm C}$, and blue lines represent $Q_{\bullet}=10^9\,{\rm C}$. \textbf{Right:} The \textit{bremsstrahlung} surface brightness profile for the case of a non-charged black hole (solid black line) and for the case of a charged black hole in the Galactic centre with the charge of $10^{7}\,{\rm C}$, $10^{8}\,{\rm C}$, and $10^{9}\,{\rm C}$, represented by dashed blue, orange, and green lines, respectively. The drops in the brightness profile with respect to the non-charged case are depicted by a sub-plot that shows the ratio of the brightness profile for the charged case with respect to the non-charged case.}
  \label{fig_brems_hole}
\end{figure*}

Let us assume that Sgr~A* is positively charged with the charge of $Q_{\bullet}$. The electrostatic potential further from the unscreened charged black hole is $\phi \approx Q_{\bullet}/(4\pi \epsilon_0 r)$. Under the assumption of the equilibrium Maxwell-Boltzmann distribution, it would yield to the charge separation and the corresponding charge particle density would vary as $n_{\rm q} \propto \exp{(-q \phi/k_{\rm B}T_{\rm p})}$ on top of the power-law (Bondi) dependency $n_{\rm q}\propto r^{-3/2}$. In Fig.~\ref{fig_brems_hole} (left panel), several number density profiles are plotted for electrons (solid lines) and protons (dashed lines) for increasing positive charge of the black hole, $Q_{\bullet}=10^{7}-10^9\,{\rm C}$, including the zero charge.

To simulate the effect of an unscreened supermassive black hole with the positive charge of $Q_{\bullet}$, we use the Abel integral to obtain the projected luminosity profile $J(R_{\rm proj})$ from the deprojected one $L_{\rm brems}(r)$, see Eq.~\eqref{eq_brems_lum},
\begin{equation}
  J(R_{\rm proj})=2\int_{R_{\rm proj}}^{R_{\rm t}} \frac{L_{\rm brems}(r)r\, \mathrm{d}r}{\sqrt{r^2-R_{\rm proj}^2}}\,,
  \label{eq_abel_intergral}
\end{equation}
where the truncation radius $R_{\rm t}$ represents the length-scale where the thermal bremsstrahlung in the Galactic centre becomes negligible. We set the truncation radius to the Bondi radius (or approximately the stagnation radius), $R_{\rm t}\approx R_{\rm B}\approx R_{\rm stag}$, see also Eqs.~\eqref{eq_Bondi_radius} and \eqref{eq_stag_Bondi}. 


The projected luminosity profile calculated using Eq.~\eqref{eq_abel_intergral} is plotted in Fig.~\ref{fig_brems_hole} (right panel). The solid line represents the case of the non-charged black hole, or a completely Debye-shielded black hole, and the dashed lines depict the cases of the supermassive black hole with the positive charge of $Q_{\bullet}=10^7-10^9\,{\rm C}$, which have progressively smaller luminosity profile than the non-charged case. A special case is the equilibrium charge $Q_{\rm eq}$, see Eq.~\ref{eq_charge_blackhole}, up to which the electron and proton number densities are comparable within a factor of a few, $n_{\rm e}\approx n_{\rm p}$, see also Fig.~\ref{fig_brems_hole} (left panel) for $Q_{\bullet}=10^7-10^8\,C$. For larger black hole charges, the drop in the bremsstrahlung profile is more prominent, specifically reaching $\sim 37\%$ of the luminosity for non-charged case at the projected radius of $R\sim 1\,{\rm arcsec}$ for $Q_{\bullet}=+10^9\,{\rm C}$. In general, the difference in the electron and proton number densities by factor $f_{\rm n}$, $n_{\rm e}=f_{\rm n} n_{\rm p}$, corresponds to the black-hole charge of,

\begin{equation}
  Q_{\bullet}(f_{\rm n})=\frac{2\pi \epsilon_0 G(m_{\rm p}-m_{\rm e})}{e}M_{\bullet}+\frac{2\pi\epsilon_0k_{\rm B}T_{\rm e,0}r_0}{e}\log{f_{\rm n}}\,,
  \label{eq_charge_factor_n}
\end{equation}
where $T_{\rm e,0}$ is the electron ($\sim$ proton) temperature at radius $r_0$. Relation~\eqref{eq_charge_factor_n} holds under the assumption of the equilibrium Maxwell-Boltzmann distribution of electrons and protons.

The calculated surface brightness in Fig.~\ref{fig_brems_hole} corresponds to the quiescent state of Sgr~A*, i.e. this profile is expected if one can remove the nonthermal variable source at the very centre. In order to satisfactorily do that, the angular resolution of X-ray instruments should be better than $\sim 0.1$ arcsec. The effect of the bremsstrahlung flattening or drop is just at (or rather beyond) the limit of what can be measured right now. Therefore this experiment and the analysis speaks for next generation X-ray telescopes that have a half and the full order of magnitude better resolving powers compared to the current situation.

\citet{2015A&A...581A..64R} construct a projected bremsstrahlung profile in their Fig. 6. At the 1$\sigma$ level, the profile shows a decrease in the brightness at radii $\lesssim 0.4''$. This is, however, still consistent within uncertainties with the flat and the slightly rising profile at the 3$\sigma$ level. Flat to slightly decreasing luminosity profile allows us to put the upper limit on the black hole charge if we assume that the charge is not screened. The projected profile inferred from Chandra observations by \citet{2015A&A...581A..64R} is consistent within the uncertainty with all profiles up to the equilibrium value of $Q\lesssim Q_{\rm eq}\approx 3.1\times 10^8\,{\rm C}$, see Fig.~\ref{fig_brems_hole} (right panel). For larger charge values, the projected profile is expected to decrease below $R=2''$, see the green dashed curve in Fig.~\ref{fig_brems_hole}, which corresponds to $Q_{\bullet}=10^{9}\,{\rm C}$. The equilibrium value of the black hole charge $Q_{\rm eq}\approx 3.1\times 10^{8}\,{\rm C}$, which was derived based on the classical mass segregation arguments, see Eq.~\eqref{eq_charge_blackhole}, corresponds to the charging/discharging length-scale of $R_{\rm charge}\approx 0.21''$ according to Eq.~\eqref{eq_r_charge}, assuming the free-fall flow. This scale is comparable to the projected radius, where the observed bremsstralung profile is consistent with the flat to decreasing flux density \citep{2015A&A...581A..64R}.

Within uncertainties, this is consistent with the constraints given by induction mechanism presented in Subsections~\ref{subsec_rotating_SMBH}, which gives an upper limit of the order of $10^{15}\,{\rm C}$. In the future, if the angular resolution of X-ray instruments is one half to one order of magnitude better,
one can distinguish the unresolved central component from the surroundings and it will be possible to model it away without assuming intrinsic physics. One could, in particular, take multiple images of the flares and model a variable point source and an
extended quiescent component. Hence, this procedure should yield a well-constrained background with a more precise brightness profile, based on which the decrease could be confirmed or excluded. 

In case the drop in the bremsstrahlung profile is confirmed on sub-arcsecond scales from Sgr~A*, one should naturally consider also other possibilities for the decrease, in particular the lower temperature due to plasma cooling and/or the decrease in the ambient gas density. However, the presence of small electric charge associated with Sgr~A* remains as an interesting possibility for both explaining the bremsstrahlung flattening as well as for testing the presence of the Debye-shell effect around supermassive black holes immersed in plasma.

\subsection{Effect of charge on ISCO of Sgr A*}

One of the important characteristics of black holes in accretion theory playing crucial role in observational constraints of black hole parameters is the location of the innermost stable circular orbit (ISCO). For a non-rotating neutral black hole, the ISCO is located at $r=3 r_S$. Rotation of black hole shifts the position of ISCO of co-rotating particles towards the horizon matching with it in the extremely rotating case $a_{\bullet} = J_{\bullet}/M_{\bullet} = 1$. The presence of the black hole charge acts in a similar way to the ISCO, shifting it towards the horizon for both neutral and charged particles \citep{2011PhRvD..83j4052P}. 
Motion of charged particle with mass $m_{\rm par}$ and charge $q_{\rm par}$ moving around non-rotating black hole with charge $Q_{\bullet}$ is restricted by the energy boundary function or the effective potential
\beq \label{energy-bound}
\frac{E_{\rm par}}{m_{\rm par} c^2} = \frac{k_1 q_{\rm par}~Q_{\bullet}}{ r } + \left[ \left(1-\frac{1}{r} + \frac{k_2 Q_{\bullet}^2}{r^2}\right) \left(1+\frac{L_{\rm par}^2}{m_{\rm par} c^2 r^2} \right) \right]^{1/2},
\eeq
where $E_{\rm par}$ and $L_{\rm par}$ are the energy and angular momentum of charged particle and $r$ is the radius given in the units of gravitational radius $r_S$. At infinity or in the absence of the fields $E_{\rm par}/(m_{\rm par}c^2) = 1$. For neutral particles at ISCO of neutral black hole $E_{\rm par}/(m_{\rm par}c^2) = \sqrt{8/9}$. The constants $k_1$ and $k_2$ are the coupling constants responsible for the interaction between charges and the gravity. For an electron around Sgr A*, the constants can be estimated as
\begin{align}
k_1 &= \frac{1}{m_e c^2 r_S^*} \approx 1.03 \times 10^{-6} \frac{{\rm s^2}}{{\rm g~cm^3}},\\
 \quad k_2 &= \frac{G}{c^4 r_S^{2*}} \approx 5.92 \times 10^{-74} \frac{{\rm s^2}}{{\rm g~cm^3}}.
\end{align}
Smallness of the constant $k_2$ representing the gravitational effect of the black hole charge imply that the effect of the black hole charge on the spacetime curvature can be neglected in most of the physically relevant cases. Indeed, the possible charge of SMBH at the Galactic centre restricted by the upper limits (see Section~\ref{limits_charge}) is not able to provide sufficient curvature of the background black hole geometry, and thus, does not influence the motion of neutral particles. However, for the motion of charged particles the effect of even small black hole charge can sufficiently shift the location of orbits, due to large values of the charge to mass ratio for elementary particles. 

\begin{figure}
\includegraphics[width=0.45\textwidth]{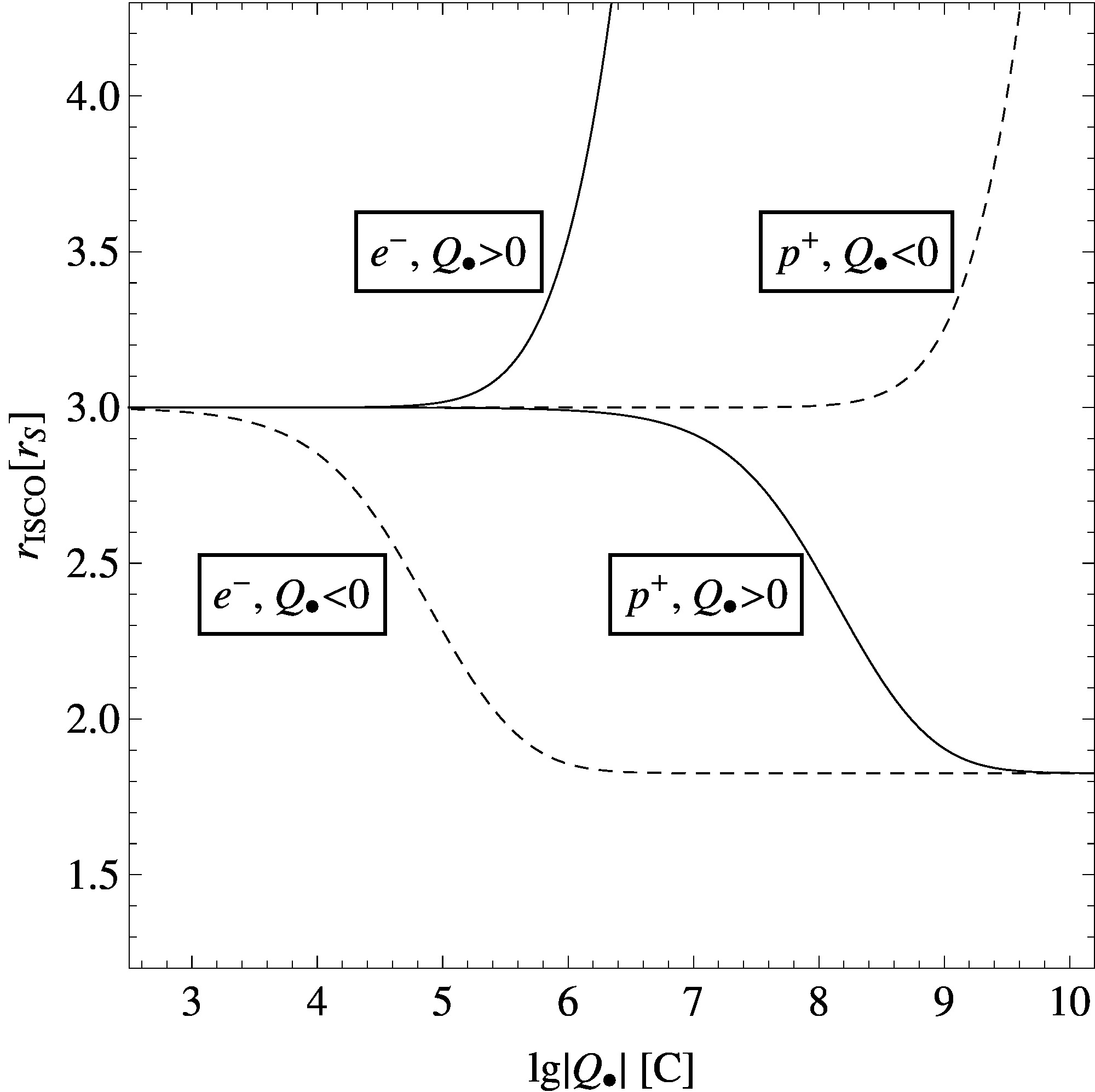}
\caption{Location of the innermost stable circular orbit of electrons $e^{-}$ and protons $p^{+}$ around Sgr A* for positive and negative configurations of the SMBH charge $Q_{\bullet}$. The radii on vertical axis are given in the units of gravitational radii of Sgr A*. }
\label{isco}
\end{figure}

The location of the ISCO of charged particles around Sgr A* as a function of the black hole charge $Q_{\bullet}$ is plotted in Fig.~\ref{isco}. As one can see from the plot, the position of the ISCO for electrons shifts from $r=3 r_S$ (corresponding to the ISCO of neutral particles) towards or outwards from black hole starting already from relatively small charges of the order of $10^3$ - $10^5$C. Thus, even a small black hole charge which does not affect the background geometry can sufficiently shift the ISCO of free electrons and protons orbiting around Sgr A*. In case of like-charges ($e^{-}$, $Q_{\bullet}<0$ or $p^{+}$, $Q_{\bullet}>0$) the ISCO can be shifted from the distance $r=3 r_S$ up to $r=1.83~r_S$, which can mimic the black hole spin with the value of $a_{\bullet} = 0.64$. This should be taken into account in the spin determination of Sgr~A*, since previous estimates are close to this value, in general above $\sim 0.4$ \citep{2010MNRAS.403L..74K,2010A&A...510A...3Z,2006A&A...460...15M}.

Particles at the ISCO can have ultrarelativistic velocities, which leads to the emision of electromagnetic radiation from the inner parts of the accretion flow. The shift of the ISCO towards the event horizon increases the gamma-factor of charges and the gravitational redshift $z = (\lambda_0 - \lambda)/\lambda_0$ of emitted photons, where  $\lambda$ and $\lambda_0$ are the wavelengths of a photon measured by local and detached observers. The shift of the ISCO radius from $r=3 r_S$ (neutral) to $r=1.83~r_S$ (charged) increases the gravitational redshift of emitted electromagnetic radiation from $z=0.225$ to $z=0.485$.

\section{Summary and Discussion}
\label{section_summary_discussion}

In this section, we summarize the constraints on the electric charge of the Galactic centre black hole. Subsequently, we look at the potential effects of the rotation on the maximum electric charge. In addition, we discuss potential non-electric origins of the black hole charge. 

\begin{table*} 
  \centering
  \caption{Summary of the constraints on the electric charge of the SMBH at the Galactic centre presented in Section~\ref{limits_charge}.}
  \resizebox{\textwidth}{!}{  
  \begin{tabular}{c|c|c|c}
  \hline
  \hline
     Process & Limit & Notes & Subsection\\
  \hline
     Mass difference between $p$ and $e$ & $Q_{\rm eq}=3.1 \times 10^8\,\left(\frac{M_{\bullet}}{4\times 10^6\,M_{\odot}} \right){\rm C}$ & stable charge & Subsec.~\ref{sec_classical_estimates}\\
     Accretion of protons   & $Q_{\rm max}^{+}=6.16\times 10^8\,\left(\frac{M_{\bullet}}{4\times 10^6\,M_{\odot}} \right)\,\rm C$    & unstable charge & Subsec.~\ref{subsec_charge_accretion}  \\
     Accretion of electrons & $ Q_{\rm max}^{-}=3.36\times 10^5\,\left(\frac{M_{\bullet}}{4\times 10^6\,M_{\odot}} \right)\,\rm C$   & unstable charge & Subsec.~\ref{subsec_charge_accretion}  \\
     Magnetic field \& SMBH rotation  & $Q_{\bullet \rm ind}^{\rm max} \lesssim 10^{15} \left( \frac{M_{\bullet}}{4 \times 10^6 M_{\odot}} \right)^2  \left( \frac{B_{\rm ext}}{10 \rm G} \right)~  \rm C$     & stable charge & Subsec.~\ref{subsec_rotating_SMBH} \\
       \hline
     Extremal SMBH & $Q_{\rm max}=6.86 \times 10^{26}\, \left(\frac{M_{\bullet}}{4\times 10^6\,M_{\odot}} \right)\sqrt{1-\tilde{a}_{\bullet}^2}\,\rm C$  & uppermost limit & Subsec.~\ref{limits_charge}\\
     \hline
  \end{tabular}
  }
  \label{tab_summary_constraints}
\end{table*}

\subsection{Charge values associated with the Galactic centre black hole}
\label{summary_charge_values}
In Section~\ref{limits_charge}, we studied limits on the electric charge of Sgr~A* based on different mechanisms, namely the accretion of charged constituents of plasma and the induction mechanism based on a rotating SMBH in the external magnetic field. We summarize these constraints in Table~\ref{tab_summary_constraints}. The charging based on accretion of protons or electrons is not stable and leads to the discharging on the discharging time-scale. However, the rotation of the SMBH within the external magnetic field is a plausible process that can result in a stable charge of the SMBH not only in the Galactic centre but in galactic nuclei in general.

All the upper limits on the electric charge in Table~\ref{tab_summary_constraints} are at least ten orders of magnitude below the maximum charge (see Eq.~\ref{eq_max_charge}) and hence the space-time metric is not affected. However, the dynamics of charged particles is significantly affected by even these small values and can be observationally tested via the change in the bremsstrahlung brightness profile.    

\subsection{Effect of rotation} 

As was already indicated by Eq.~\eqref{eq_max_gen_nodim}, the rotation of the SMBH does effect the value of the maximum allowed charge for a black hole. For the maximum rotation, $\tilde{a}_{\bullet}=1$, the maximum charge vanishes completely. However, the dependence of $Q_{\rm max}^{\rm rot}$ on $a_{\bullet}$ is only prominent for large spins, see also the left panel of Fig.~\ref{fig_rel_correction}. In astrophysical relevant systems, the maximum rotation parameter is $a_{\bullet}^{\rm max}=0.998$ \citep{1974ApJ...191..507T}, which results in $Q_{\rm max}^{\rm rot}\approx 0.06\,Q_{\rm max}^{\rm norot}=4.3\times 10^{25}\,C$. This is still about ten orders of magnitude larger than the constraints analysed in Section~\ref{limits_charge}. For Sgr~A*, the spin is estimated to be even smaller, $a_{\bullet} \sim 0.5$ \citep{2006A&A...460...15M,2010A&A...510A...3Z}. Therefore, the previous analysis is valid also for the case when Sgr~A* has a significant spin.

\subsection{Non-electric origin of the black hole charge}

The black-hole charge can also be of a non-electric origin \citep{2014PhRvD..90f2007Z}, namely a tidal charge induced by an extra dimension in the Randall-Sundrum (RS) braneworld solutions or 5D warped geometry theory \citep{1999PhRvL..83.3370R,2010PhRvD..81l3011B,2011CQGra..28k4003B}, in which the observable Universe is a $(3+1)$-brane (domain wall) that includes the standard matter fields and the gravity field propagates further to higher dimensions.  The RS solution yields the 4D Einstein gravity in the low energy regime. However, in the high energy regime deviations from the Einstein solution appear, in particular in the early universe and the vicinity of compact objects \citep{2009GReGr..41.1305S}. Both the high-energy (local) and the bulk stress (non-local) affects the matching problem on the brane in comparison with Einstein solutions. In particular, the matching does not result in the exterior Schwarzschild metric for spherical bodies in general \citep{2008CQGra..25v5016K}.

A class of the RS brane black-hole solutions is obtained by solving the effective gravitational field equations given the spherically symmetric metric on the $(3+1)$-brane \citep[see][for a review]{2004LRR.....7....7M}. These black holes are characterized by Reissner-Nordström-like static metric which has the non-electric charge $b$ instead of the standard $Q^2$. This charge characterizes the stresses induced by the Weyl curvature tensor of the bulk space, i.e. 5D graviton stresses that effectively act like tides. Thus, the parameter $b$ is often referred to as the tidal charge and can be both negative and positive. Detailed studies of the optical phenomena associated with brany black holes, in particular quasiperiodic oscillations (QPOs) that are of an astrophysical relevance, were performed in several studies \citep{2009IJMPD..18..983S,2009GReGr..41.1305S}. 

As pointed by \citet{2014PhRvD..90f2007Z}, the tidal charge could be tested by detecting the black hole shadow. While the electric charge causes the shadow to shrink, which is only noticeable for nearly extremal charges (see Subsection~\ref{effect_shadow}), the tidal charge could act in the opposite sense -- enlarging the shadow size. Thus, if a noticeable change in the shadow size is detected, it would most likely be caused by non-electric tidal charge since the electric charge is expected to have negligible effects on the metric as we showed in Section~\ref{limits_charge} and summarized in Subsection~\ref{summary_charge_values}.

\subsection{Comparison with previous studies}

\citet{2014PhRvD..90f2007Z} uses an argument that the measured core size of Sgr~A* of $\sim 40\,{\rm \mu as}$ \citep{2008Natur.455...78D,2011ApJ...727L..36F} is more consistent with the Reissner-Nordström black hole with the charge close to the extremal value of $Q_{\rm max}^{\rm norot}$ than with the Schwarzschild black hole whose shadow is expected to be $\sim 53\,{\rm \mu as}$. However, one needs to stress that the core size does not directly express the shadow size at all since it does not have to be centered at the black hole -- it can, for instance, be caused by a Doppler-boosted accretion flow or the jet-launching site \citep{2012MNRAS.421.1517D,2017FoPh...47..553E}. Hence, the charge constraint given by \citet{2014PhRvD..90f2007Z} is uncertain at this point and should be further tested when the analysis of the observations by the Event Horizon Telescope\footnote{\url{http://eventhorizontelescope.org/}} is available.

\citet{2012GReGr..44.1753I} gives the following estimate on the charge of Sgr~A* based on the geodesic trajectories of the orbital motion \citep{2001CeMDA..79..135C}, in particular using S2 star orbit, $Q_{\bullet}\lesssim 3.6 \times 10^{27}\,{\rm C}$, leaving space for the charge larger than an extremal value. 

In comparison with \citet{2012GReGr..44.1753I} and \citet{2014PhRvD..90f2007Z}, we take into account the presence of plasma and the magnetic field in the vicinity of Sgr~A*, which leads to tighter constraints and significantly smaller values, see Table~\ref{tab_summary_constraints}. Moreover, our suggested test based on the bremsstrahlung brightness profile, see Subsec.~\ref{subsec_brems_hole}, is significantly more sensitive to smaller charge than the shadow size or stellar trajectories.


\section{Conclusions}
\label{section_conclusions}

We performed analytical calculations to find out if the supermassive black hole at the Galactic centre can get charged and what the realistic values of its charge are. Based on the classical estimates and total amount of the charge in the sphere of influence of the black hole, we expect that the black hole can acquire a small, transient positive charge of $\lesssim 10^9\,{\rm C}$, which does not have an influence on the spacetime metric. Based on the general relativistic calculations, we further explore the induced charge based on the rotating black hole that is immersed in the external magnetic field. Such a configuration is in general expected in almost all galactic nuclei. If the black hole spin axis is approximately aligned with the external magnetic field, we again expect that the induced charge is positive, with the uppermost limit of $Q_{\bullet}\lesssim 10^{15}\,{\rm C}$. Although the spacetime metric is not influenced significantly by electric charge within the limits we found, even such a small charge can significantly influence the typical viscous timescales for protons and electrons as well as an infall of small charged particles (dust particles). Most importantly, for like charges of test particles and Sgr~A*, the ISCO shifts significantly in comparison with the no-charge case even for a small charge of the order of $10^{6}$--$10^{10}\,{\rm C}$, which effectively mimics the black hole spin of $a_{\bullet}\sim 0.6$. This effect should be taken into account in the future numerical calculations as well as the analysis of observation data.

 We also revisited observational tests of the presence of the charge for the Galactic centre black hole. The black shadow size, which was proposed previously, is only sensitive for large values of the charge, close to an extremal value, which are unrealistic as we showed. We propose a new test based on the observed surface brightness profile of the thermal bremsstrahlung inside the innermost $10^5$ Schwarzschild radii, which is the region that coincides with the S cluster. Within this range, a flattening and a decrease in the bremsstrahlung surface brightness is expected to occur due to the presence of the charged, unshielded black hole starting with about twenty orders of magnitude smaller values than the extremal case. Since the Chandra X-ray observations with the angular resolution of $0.5''$ did detect a weak indication of the drop in the brightness profile at $R_{\rm proj}\lesssim 0.4''$, it puts an observational upper limit on the charge $Q_{\rm SgrA*}\lesssim 3\times 10^{8}\,{\rm C}$.

\section*{Acknowledgements}

 We thank Vladim\'ir Karas and Zden\v{e}k Stuchlík for very useful comments on the manuscript. We are especially grateful to Elaheh Hosseini for checking the calculations. We received funding from the European Union Seventh Framework Program (FP7/2013-2017) under grant
agreement no 312789 - Strong gravity: Probing Strong
Gravity by Black Holes Across the Range of Masses.
This work was supported in part by the Deutsche
Forschungsgemeinschaft (DFG) via the Cologne Bonn
Graduate School (BCGS), the Max Planck Society
through the International Max Planck Research School
(IMPRS) for Astronomy and Astrophysics, as well as
special funds through the University of Cologne and
SFB 956 Conditions and Impact of Star Formation.
M. Zaja\v{c}ek is a member of the International Max Planck Research School at the Universities of Cologne and Bonn. Arman Tursunov acknowledges the Czech Science Foundation Grant No. 16-03564Y and the Silesian University in Opava Grant No. SGS/14/2016.













\bsp	
\label{lastpage}
\end{document}